\documentclass{article}

\usepackage[english]{babel}

\usepackage{amsmath}
\usepackage{graphicx}
\usepackage[colorlinks=true, allcolors=blue]{hyperref}
\usepackage{pdfpages}
\usepackage{xurl}
\usepackage[utf8]{inputenc}
\usepackage{arxiv}
\usepackage{orcidlink}
\newcommand{\mycomment}[1]{}

\author{
  Montassar Naghmouchi \orcidlink{0000-0003-3467-7514}\\
  SAMOVAR, Télécom SudParis, Institut Polytechnique de Paris \\
  \texttt{montassar-bellah\_naghmouchi@telecom-sudparis.eu} 
    \\
    \And
  Maryline Laurent \orcidlink{0000-0002-7256-3721}\\
  SAMOVAR, Télécom SudParis, Institut Polytechnique de Paris \\
  \texttt{maryline.laurent@telecom-sudparis.eu} 
}

\title{A Systematic Review and Layered Framework for Privacy-by-Design in Self-Sovereign Identity Systems}

\begin{document}

\maketitle

\begin{abstract}
The use of Self-Sovereign Identity (SSI) systems for digital identity management is gaining traction and interest. Countries such as Bhutan have already implemented an SSI infrastructure to manage the identity of their citizens. The EU, thanks to the revised eIDAS regulation, is opening the door for SSI vendors to develop SSI systems for the planned EU digital identity wallet. These developments, which fall within the sovereign domain, raise questions about individual privacy.

The design of SSI systems is complex, often characterized by a large number of components and architectural choices because the current SSI communities differ on how to create identifiers, how to build and present credentials, and even how to design a user wallet. SSI stacks developed by different organizations provide different privacy features for different privacy needs.

This paper performs a systematic mapping and review of SSI components and technologies into a novel four-layer privacy framework to address the design complexity of SSI systems. Based on this review, we provide an accompanying Design Assistance Dashboard (DAD).
The DAD shows the interdependencies between SSI components in different layers, and maps these components to different privacy requirements and considerations, even providing a simple privacy class for each component.
\mycomment{
The purpose of this article is to help SSI solution designers make informed choices to ensure that the designed solution is privacy-friendly. The observation is that the range of possible solutions is very broad, from DID and DID resolution methods to verifiable credential types, publicly available information (e.g. in a blockchain), type of infrastructure, etc. As a result, the article proposes (1) to group the elementary building blocks of a SSI system into 5 structuring layers, (2) to analyze for each layer the privacy implications of using the chosen building block, and (3) to provide a design assistance dashboard that gives the complete picture of the SSI, and shows the interdependencies between architectural choices and technical building blocks, allowing designers to make informed choices and graphically achieve a SSI solution that meets their need for privacy.\\}

\end{abstract}
\newpage

\section{Introduction}
\label{section 1}

As clearly presented in NIST's white paper \cite{NIST-whitepaper}, the ID management systems have been evolving from \textbf{siloed identity} systems, to  \textbf{federated identity}, \textbf{user-centric identity} and more recently \textbf{SSI} (for  \textbf{Self-Sovereign Identity}). Each model has a different view on the degree of privacy needed. SSI is the most privacy-cautious model. It gives users complete control, autonomy and ownership of their identifiers and enables them to identify and authenticate themselves with any other entity without referring to an Identity Provider (IdP) or trust authority. SSI also enables the creation and exchange of credentials that are cryptographically verifiable in terms of authorship, ownership and revocation. SSI relies on a wallet application which is used by its owner to generate, control and own their identity management data, like identifiers, cryptographic keys, tokens, credentials and others. It relies also on a decentralized infrastructure which publishes public identifiers, public keys and metadata, and records identity-related transactions.

A number of initiatives have emerged in recent years with the SSI philosophy in the background. With the release of the FIDO 2.0 framework in 2015, \cite{FIDO-W3C} by the FIDO alliance, the World Wide Web Consortium (W3C, \cite{W3C}) and the Decentralized Identity Foundation (DIF,  \cite{DIF}) - together with a consortium of other partners - developed the vision of an identity management model with its own standards that can operate without a central trust authority, and can go even further by being truly open and not owned by any single entity or group of entities. In 2023, Bhutan became the first country in the world to establish a SSI infrastructure to manage the identities of its citizens \cite{Bhutan-SSI}. Countries such as South-Korea \cite{SA-SSI}, which has created a consortium for SSI solutions, and the EU which is currently working on SSI systems for EU cross-border identity for EU member states, are other examples of the traction SSI is gaining at scale.
Ahead of schedule with the revision of the eIDAS Regulation \cite{eIDAS}, the European Commission (EC) has funded several projects such as the European SSI Framework (ESSIF) \cite{ESSIF} and the European Blockchain Services Infrastructure (EBSI) \cite{EBSI} for several use cases, including the European Digital Identity. The EC has recently funded four large scale pilots to build and test prototypes of an EU digital identity wallet, and plans to equip 80\% of European citizens with such a wallet by 2030.

Two observations can be made about SSI systems and their use as digital identity systems, especially on a large scale:
\begin{enumerate}
    \item Although the SSI community is converging on standards such as DID (Decentralized Identifier), VC (Verifiable Credential) and others, in terms of design and implementation, SSI solutions could not be more different.
    
    \item Maintaining privacy in identity management systems is a challenging issue - even more so as society moves legal identities into the digital age - to avoid falling into a society of facilitated and enhanced surveillance.
\end{enumerate}

Based on these observations, this paper is intended to help gain a deep understanding of the SSI building blocks (hereafter referred to as components) and their functional interdependencies, and to understand the privacy implications of possible architectural choices.

The contributions of the paper can be summarized as follows: 
\begin{enumerate}
    \item A deep survey and review of SSI ecosystems. The major SSI components are mapped into a layered framework that associates technological components with privacy requirements and considerations.
    \item A Design Assistance Dashboard (DAD), which is a graphical representation of the framework, known as a design graph. It consists of displaying a set of interconnected components belonging to different layers. It helps to show dependencies between SSI layers and components, and conflicts between components. The design graph can be integrated into the computer-aided design of SSI systems. By providing a quick view of SSI systems, it enables the SSI system designer to quickly analyze SSI system proposals and make an informed choice for a SSI solution that satisfies the need for privacy.
\end{enumerate}

The remainder of this paper is organized as follows. 
In Section \ref{methods}, we present the research methodology and describe how the SSI ecosystems were mapped and reviewed. We also explain how the privacy-oriented framework and DAD were constructed.
Section \ref{section 2} provides an overview of SSI systems, introduces our four layer privacy model of SSI systems.
Section \ref{RelatedWork} presents the related work on the SSI layered models and some privacy related work, highlighting the lack of granularity in the existing work to analyze the privacy of SSI systems.
The next four sections review the four layers and their components, highlighting the diversity of possible approaches and the privacy implications of each component. \ref{section 3} analyzes and compares the possible infrastructures to be used with SSI, the base layer of SSI systems. Section \ref{section 4} addresses the identifiers and cryptographic material needed for identification and authentication in SSI systems, focusing on the most prominent and statistically used SSI identifiers.
Section \ref{section 5} examines SSI credentials, the critical presentation phase of SSI credentials by their owners, and revocation, which is a mandatory feature but often conflicts with the need for privacy.
In Section \ref{section 6}, the upper layer of the SSI systems that make up the wallet application is presented with a novel classification of wallets based on their architectures.
After presenting our original Design Assistance Dashboard (DAD) accompanied by a step-by-step application guide in Section \ref{section 7}, Section \ref{section 8} is dedicated to discussion, challenges and future works.

\section{Methodology}
\label{methods}
In order to construct a framework that is comprehensive in its review and analysis, yet practical in its application, we adopted a two-phase methodology.
In the first phase, we conducted a thorough technology and literature review to map and analyze SSI components and the ecosystem. In the second phase, we synthesized our findings to create our proposed layered privacy framework and the accompanying Design Assistance Dashboard (DAD).

\textbf{Phase 1: Systematic Review} \\
This phase is crucial for deconstructing the complex SSI landscape, which ranges from new standards (both accepted and still in draft form) to SSI working groups under different organisations, literature and academic proposals closely related to SSI, and industrial products such as SSI wallets provided by SSI vendors. We have a multi-source review process: \\

1. \textit{Analysis of standards and specifications}: An in-depth analysis of current SSI standards and specifications, ranging from adopted standards to simple editor drafts. Technical documentation from the W3C on core data models such as DIDs, VCs and VPs was studied and analysed extensively. Protocols from DIF, such as DID-Comm (v1 and v2), and early DID-Auth proposals from RWOT (Rebooting the Web of Trust), were also studied. This enabled us to develop a foundational understanding of how SSI components are designed to function.
\\
2. \textit{Review of open-source implementations}: The gap between specifications and practice is bridged by reviewing the architectures and documentation of influential open-source SSI stacks and projects. The most well-known example is the Linux Foundation's Hyperledger ecosystem, which comprises Hyperledger Indy, Aries and AnonCreds. This in-depth analysis sheds light on how theoretical standards are implemented in real-world software stacks and the trade-offs between security, privacy, and performance.
\\
3. \textit{Assessment of real-world products and industry data}: To ensure that our framework reflects currently used technologies, we prioritised the SSI components with the widest adoption. This selection was guided by industry data, such as the CheckD survey of 37 SSI vendors (cited in {checkD-survey}), which provides quantitative data on the market share of competing DID methods and VC schema. We also closely studied some SSI wallets and even exchanged views with the technical teams behind them to understand the practical challenges and implementation patterns.
\\
\\
\textbf{Phase 2: Framework Synthesis and Development}\\
In this phase we synthesized the findings from phase 1 in two steps: \\
1. \textit{Construction of the layered privacy framework}: We categorized the identified components based on their functional role in a typical SSI interaction (e.g., identification, attestation and user interaction). This bottom-up, functional grouping naturally led to the development of our proposed four-layer model: Infrastructure, Identifiers \& Cryptographic Material, Credentials \& Presentations, and Wallet Applications. While there are similar models (cf. Section \ref{RelatedWork}), we take it a step further by identifying privacy requirements from different sources, such as regulation, industry standards, community discussions. We then map these privacy properties to the four SSI layers. The result is a layered privacy model in which we can analyze each layer and its components from a privacy perspective with defined goals and requirements.
\\
2. \textit{DAD creation}: DAD builds on the previous layered model, adding two extra steps. Firstly, DAD identifies the dependencies and architectural possibilities between the different components. The second step involves attributing a privacy indicator to each component, which allows the privacy level to be visualized using a simple colour code. Please note that this privacy indicator is a simplified privacy rating based on our analysis and review provided for each component in Phase 1.
DAD is a graphical representation of our SSI review process. It allows users to analyze and design SSI systems by following a simple flowchart on a graphical interface.
\\

This two-phase methodology ensures that our review is not just theoretical, but has a practical application in the form of the DAD.
\section{SSI Systems and Privacy}
\label{section 2}
The SSI model removes the traditional roles found in legacy identity management models such as the IdP. SSI has a different vision of identity where no registration authorities are required, favoring decentralized infrastructures and autonomous identifier generation and registration methods in order to avoid depending on an IdP. Moreover, SSI retains the user-centric model vision where the user is in the middle of interactions, receiving credentials from issuers, and presenting them to verifiers. Since it was the job of the IdP to provide privacy for users by only sharing what needs to be shared with Relying Parties (RPs) and protecting the identity management data of all users, in SSI this responsibility is shifted to the holder (user), which means that privacy considerations and requirements for SSI are different from other identity management models. 

\begin{figure}[h]
    \centering
    \includegraphics[width=0.8\textwidth]{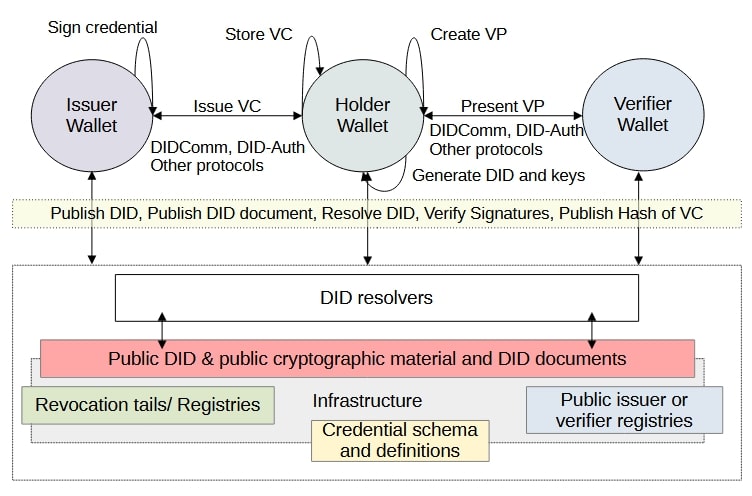}
    \caption{The SSI system model and architecture.}
    \label{fig:1}
\end{figure}

\subsection{SSI Systems}
\label{section 2.1}
As shown in Figure \ref{fig:1}, three (3) main roles can be found in a SSI system. Each role is autonomous in generating their identifiers and cryptographic materials (private keys and public keys, secrets, commitments) needed to create, exchange and verify credentials. This is empowered by a \textbf{wallet application} and an \textbf{infrastructure}. 

There are three main actors in an SSI system:
\begin{itemize}
    \item \textbf{Issuer} creates credentials for holders. 
    After receiving a holder's autonomously generated identifier, the issuer creates a set of claims for that holder, grouping them together with that identifier in what is known as a credential. The issuer is responsible for verifying statements about the holder and is generally trusted to provide trusted credentials.
    \item \textbf{Holder} is at the center of the SSI model. The holder requests credentials from the issuers (the issuance phase), stores them in their wallet, and presents proofs of credentials and ownership of the credentials to the verifiers when needed (the presentation phase).
    \item \textbf{Verifier} requests proofs from holders and checks their validity before providing a service or access.
\end{itemize}

These roles are interchangeable, meaning that, depending on the context, a user may well take on one or other of these roles. 
The interaction between different stakeholders is made possible by several SSI building blocks that support the generation, presentation, and verification of credentials. The layered representation of these defined building blocks is illustrated in Figure \ref{fig:new} (cf. Section \ref{RelatedWork} for arguments in favor of this model) and refers to the following layers: 

\begin{enumerate}
\item \textbf{Infrastructure} (cf. Section \ref{section 3}): Decentralized infrastructures are especially advocated to use with SSI systems to satisfy the principles of ownership and user control since we can argue that a private infrastructure owned and controlled by a single entity or a limited number of entities can not possibly satisfy self-sovereignty for users. Blockchains, DLTs in general and even decentralized web-based infrastructures are all SSI compatible. The infrastructure is used to enable certain other components like identifiers and public keys (as a Decentralized Public Key Infrastructure DPKI) but also for certain functions such as storage and even computing. 

\item \textbf{Identifiers \& cryptographic material} (cf. Section \ref{section 4}): Identifiers and cryptographic material like public keys are used to identify SSI users, no matter the role they assume. Authenticating the users via their identifiers provides the two basic functionalities of identity management: identification and authentication. While other models rely on IdPs to create identifiers for system users and to provide authentication mechanisms (like OpenID Connect and authentication servers), SSI relies on identifiers that do not require registration authorities or IdPs. The most common identifiers used in SSI solutions follow the DID standard, created by W3C. These DIDs  are created by their owner, they are linked to a public key that is associated to a private key. The ownership of the identifier is verified by a challenge/response (cf. Section \ref{section 6.4.2}) where a message is encrypted using the public key of the DID owner that can be publicly available or privately transmitted to concerned entities in the case of identifiers intended for private use. Public keys and other data related to the authentication of a DID, how to communicate with the DID owner or controller – called endpoints – can be specified in a \textbf{DID document}. Mapping the DID identifier to a DID document is essential and it is the first step in each SSI interaction. This operation is called \textbf{“DID resolution”} and requires a special component called a \textbf{"DID resolver"}. 
The simplest DID document is a public key that can be derived from the identifier itself. Depending on the  \textbf{DID method}, resolvers use different logic but they all follow the same resolution algorithm \cite{DID-resolution} (cf. Section \ref{section 4}).
Public identifiers are typically found on the infrastructure, which is generally the case for the identifiers of discoverable users like issuers and verifiers. We also add a sub-layer \textbf{identifier resolution} that operates between the \textbf{identifiers \& cryptographic material} layer and the \textbf{Infrastructure} layer for better logical separation between SSI layers.

  \item \textbf{Credentials \& Presentations} (cf. Section \ref{section 5}):  
  \textbf{Credentials} are a set of attributes, put together in a given data model (notably JSON and JSON-LD), they contain a proof that makes them verifiable in terms of authorship (issuer-side) and ownership (holder-side). Generally, credentials are signed with keys that are related to DIDs created by the users. A credential is associated to another external document called \textbf{credential schema} that describes the attributes, expected properties and formats, their types, etc. The document is in the form of a template/blueprint to create credential of that type. A \textbf{credential definition} is an instance of a credential schema which is created by an issuer to state that they intend to issue credentials of that specific type. It contains the issuer’s identifiers, cryptographic keys or methods used to sign and verify the credential which is issued according to this given scheme (Assertion methods). 
    These two documents are mentioned in each credential. JSON-LD uses an Internationalized Resource Identifier (\textbf{IRI}) to point to these documents as resources. Other approaches publish them on a Blockchain like Hyperledger Indy or even in the credential metadata themselves, which leads to bigger credentials (cf. Section \ref{section 5.1}).  

    The \textbf{Presentation} phase consists of the holder presenting the requested attributes to a verifier, either by presenting the whole credentials, or by selecting and combining the attributes of the different credentials to be presented (cf. Section \ref{section 5.2}).

   \item \textbf{Edge Agent \& Cloud Agent (Wallet Application)} (cf. Section \ref{section 6}): 
   The higher layer of the SSI includes communication protocols between peers, the storage and exchange of credentials and presentations, and the signing or verification of credentials. All of these functionalities are handled via a wallet application. The wallet application enables users to generate their identifiers and cryptographic keys, and publish them – if needed – on the infrastructure. It also enables users to authenticate themselves in peer-to-peer, meaning that the wallet is capable of DID resolution. A wallet is generally composed of a cloud agent that runs on a cloud as an online service and an edge agent that runs on the user’s device.   
\end{enumerate}

Although we try to separate SSI systems into major components and layers, it is important to understand that there is a high dependency between layers in this layered description of SSI. 

Refer to  \textbf{Appendix A} for a complete breakdown of the SSI into architectural and technological components and layers.

\subsection{Privacy Requirements in SSI}
\label{section 2.2}

The following privacy properties are generally desirable for digital identity management systems and especially those that follow the SSI model:
\begin{itemize}
\item \textbf{Pseudonymity}: The holder can generate multiple identifiers and choose which identifier to use for each verifier. This means that the holder may have different identifiers for different relationships with other actors, which is satisfied for example using \textbf{pairwise DIDs} \cite{pairDID}.
\item \textbf{Multi-Show Unlinkability}: Several credential presentations derived from the same original credential and transmitted over several sessions cannot be linked by the verifiers.
This can only be achieved by using non-correlatable identifiers that can be changed at each interaction, such as \textbf{link secrets} and by implementing signature schemes or signature proofs that can be randomized at each presentation, such as Camenisch-Lysyanskaya Signatures \cite{CL-sig} or BBS+ signatures \cite{BBS}. 

\item \textbf{Issue-Show Unlinkability}: Any information collected at the time of credential issuance cannot be used subsequently to establish a link between the credential presentation and the original credential.

\item \textbf{Non-Correlation}: Non-correlation is the inability of any actor, including curious actors on the infrastructure for example or providing a service like DNS servers, to learn about the activity of the holder, by linking their different identifiers together, or by linking an activity to the specific holder. Non-correlation is desired in multiple SSI components and layers. In the Credentials \& Presentations layer, it is referred to as multi-show unlinkability or issue-show unlinkability because verifiers and issuers handle specific holder information, i.e. credentials.

\item \textbf{Selective Disclosure}: The holder is able to select the information to disclose to the verifiers during the presentation phase. This can be the attributes selected from his credentials. The other case is based on \textbf{Zero-Knowledge Proof (ZKP)} \cite{ZKP} which enables the holder to prove certain assertions or predicates about a claim without revealing the real value of the claim itself. The holder  can prove that he is in possession of a certain credential without revealing the credential itself. An example is proving that the holder is a student, only by showing a ZKP of possession of an enrollment certificate credential at an educational institution, or showing that the holder's age is over 18 using a ZKP from the date of birth found on a birth certificate credential.

\item \textbf{Transparency}: Transparency is fundamental for privacy and data protection, since it guarantees that the data subject is aware of how their data are being processed and where it is stored, etc. Transparency enables data to be controlled, which is why it is an important property that any digital identity management infrastructure must satisfy. Transparency however must not come at the expense of data privacy and the confidentiality of certain transactions.

\item \textbf{Non-traceability}: Non-traceability means that a curious actor is not able to trace the activities of a certain user back to them. 

This is a desired property of a privacy-friendly identity management system, such as a SSI, but it conflicts with the need for accountability, which requires that only a trusted actor be able to trace activities when needed. Non-traceability may be associated with more than one SSI layer. This is the case of the Architecture and Reference Framework \cite{ARF} (Draft 0.0 of October 2022) specified by the eIDAS Expert Group, which considers \textbf{non-traceability of wallet activity} as a privacy requirement to be met by the wallet provider.

\item \textbf{Confidentiality of wallet communication}: Credentials or messages exchanged between two actors via their two wallets must remain confidential.
The confidentiality of the communication between wallets is an important privacy requirement to ensure the security of the SSI interactions.

\end{itemize}

\begin{figure}[h]
    \centering
    \includegraphics[width=0.8\textwidth]{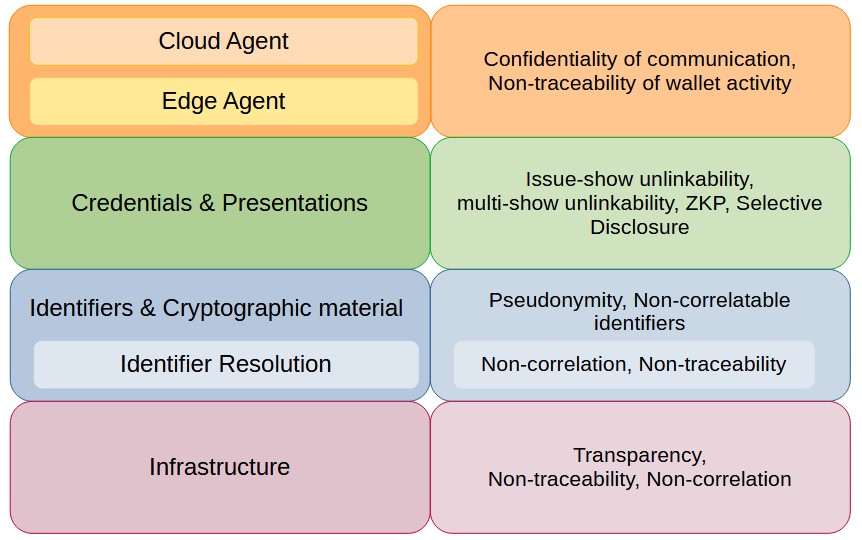}
    \caption{Mapping privacy properties to each SSI layer.}
    \label{fig:new}
\end{figure}

Referring to the SSI layered model, it is important to understand that the slightest lack of privacy in one layer calls into question the privacy of the whole. In the same vein, a privacy requirement satisfied in one given layer becomes an inherited feature for other layers. For example, if the \textbf{Identifiers and Cryptographic material} layer satisfies \textbf{\textit{non-correlation}} or \textbf{\textit{pseudonymity}} by providing identifiers like \textbf{link secrets}\footnote{Link secrets are similar to \textit{master secret keys} used in IDEMIX \cite{IDEMIX}. They enable Anonymous Credential systems to bet set up, which are the basis for widely used credentials such as AnonCreds.}, the \textbf{Credentials and Presentations} layer will be able to build upon it, using privacy-preserving signature schemes, to provide anonymous credential systems, which will then satisfy additional privacy requirements, such as selective disclosure with ZKP and credential issue/multi show unlinkability. This requires building up the design in a coherent way in which every layer makes maximum use of privacy mechanisms implemented in other layers. Figure \ref{fig:new} also maps privacy requirements to SSI layers.

\mycomment{SSI systems must satisfy these properties according to the level of authentication needed by each verifier. For instance, proving to an online service provider that a holder is over 18 can be done by showing a government issued ID (like taking a passport photo) that contains the date of birth. However, this means that under several data protection laws (like GDPR), the online service provider is now a collector and/or a processor of personal data. Instead, the selective disclosure property with ZKP and specific anonymous credential, e.g. AnonCreds, can be implemented so that the holder can prove he or she is over 18 thanks to a government-issued credential, while not revealing any other information.}

\section{Related Work}
\label{RelatedWork}
Representing SSI systems as a stack or a layered model is not new. Based on the Trust over IP (ToIP) model \cite{ToIP} which encompasses SSI systems with their governance and related trust protocols, other models and mappings of the SSI components can be found in \cite{s22155641}, \cite{mapping}, \cite{Yildiz} and \cite{layering-diagram}. In \cite{mapping}, H. Yildiz provides a mapping of different SSI standards and components. Organized by layers, components are also assigned a level of maturity ranging from stable to experimental.

Our layered dissection of SSI presented in Section \ref{section 2.1} differs from other related work. Contrary to models that group the identifiers and the cryptographic material together with verifiable registries (like in \cite{layering-diagram}, DIF documentation), e.g. blockchains or databases containing public keys, our model dedicates an independent layer "Identifiers \& Cryptographic materials" which includes an "Identifier Resolution" sub-layer. This sub-layer, which manages the identifier resolution, acts as an intermediary layer between the infrastructure layer integrating the useful verifiable registries and the identifiers layer. We are also different from models where the agent layer and the application layer are two separate layers, like in \cite{Yildiz}, or where the infrastructure is combined with the agents, like in \cite{layering-diagram}. We consider the wallet application as the application layer that is composed of two sub-layers: the edge agent, which is the part of the application installed on the user's device and the cloud agent, which enables communication between edge agents in cases where the architecture is based on such server-like agents (not the case in P2P networks for example).

The objective of our model is different from other existing models and is to provide an in-depth privacy analysis for SSI systems, which requires mapping privacy properties to different SSI components and layers. Most other models focus on interoperability, and researchers focus on how to make different SSI components and layers interoperable with existing systems and between SSI systems themselves. 

Other related works deal with certain privacy issues in SSI. In \cite{9297357}, the authors study the privacy shortages of credentials in Hyperledger Indy and propose a model of attribute sensitivity score that assesses different attributes found in verifiable credentials by attributing them a score of 0.0 to 1.0, 1.0 being the highest sensitivity score for an attribute to be shared. This score model is integrated into the cloud agent (Hyperledger Aries here) (cf. Section \ref{section 6.2}) to enable it to take into account the sensitivity of the attributes presented (here Hyperledger Aries) (see Section \ref{section 5.1.2}) to a certain verifier, which can be used to raise privacy awareness and improve the overall privacy of SSI credentials. 
In \cite{DID-Privacy, VCPrivacy, VC-Privacy}, the W3C provides some Privacy Considerations related to DIDs, and VC data models 1.0 and 2.0, however, these considerations are non-normative, remain very generic and lack any notable analysis of existing SSI technologies and systems.
In \cite{10.1007}, the objective is to derive SSI credentials from existing eID systems, and to build privacy-preserving credentials that can be easily accepted by established systems, e.g. governments and banks, and that implement the selective disclosure principle thanks to ZKP.

Compared to the literature, this paper provides a deeper and broader assessment of which privacy properties a SSI system can ensure at different layers. It gives a rich comparative analysis between different SSI components in each layer. Since the objective of the paper is to focus on design, we focus on the most commonly used SSI components and provide a mapping between these components and the privacy properties that they may need to satisfy.

\section{SS Infrastructure}
\label{section 3}

An SSI infrastructure is used for one or more of the following reasons: 
\begin{itemize}
\item As a public key infrastructure (PKI) that manages public keys (published on their own or in a DID document). If the infrastructure is decentralized, it is referred to as a Decentralized PKI (DPKI). 
\item As a registry for public identifiers and public DID documents that contain – among others – the endpoints used to communicate with the owner and the authentication methods.
\item To publish hashes of credentials, cryptographic commitments, timestamps, etc.
\item To publish revocation lists, revocation tails or revocation information about credentials.
\item To publish credential schema and definitions.
\item To publish trusted lists of issuers, verifiers, and public stakeholders.
\end{itemize}

The infrastructure layer can use blockchain technology and/or web-based technology. 
Most SSI solutions rely on blockchain infrastructures in order to satisfy the SSI conditions of autonomy and ownership \cite{10.1145/3486622.3493917}. It is also shown that non-blockchain infrastructures can also satisfy these conditions, so while blockchain is not explicitly needed for SSI systems, it is a good foundation to build upon \cite{DBLP:journals/corr/abs-1904-12816}.

Both blockchain and web-based technology can operate as an open or closed infrastructure, which affects the properties that can be expected of them. 
Open infrastructure, which are operated by any entity on a voluntary basis, are transparent and accessible to everyone. They are harder to protect from curious actors which may attempt to undermine the privacy of users’ activities (non-correlation  property) and from malicious actors, which can try to corrupt the working of the infrastructure itself. 
A contrario, closed infrastructures that are owned and controlled by a few entities, usually have the advantage of providing a higher level of trust compared to open infrastructures. However, they raise more concerns about privacy, possible illegal processing of user data, lack of transparency and lack of user autonomy.

Let us explore these ideas in more detail by looking at the elements stored in the infrastructure. These elements can be used to verify the identity of different subjects and the authenticity of the credentials they present. Public keys, credential schemas and definitions, revocation information and cryptographic commitments are essential for the verification of credentials and the identity of their issuers and holders. This means that at the stage of reading or retrieving data from the infrastructure for credential verification, a public open infrastructure - such as a public blockchain - would require virtually no interaction between the verifier and the issuer or any other third party, as the necessary verification data are available on an open infrastructure. However, if such data are placed on a permissioned infrastructure, such as a permissioned blockchain or a web-based infrastructure with access control, then access to data are subject to authorization by the issuer or other entity that controls access to the data. SSI can support both visions of credential verification where it is either open and free and the issuers are not involved in the verification phase (using open infrastructures) or where the issuers (or an entity acting on their behalf) are involved. While the first model is more in line with the SSI principles of autonomy, privacy and non-correlation of credential verification, the second model offers monetization of credential verification, where the issuer can make money from the credentials they issue by having a verification fee paid directly to them each time one of their credentials is verified. Some SSI vendors, such as Dock \cite{Dock}, offer "paid credential verification" as an option. If the issuer is needed for the verification of the credentials (e.g. by providing the material needed for verification on their web server and controlling the verifiers who access it, by allowing access to private channels on the blockchain or the blockchain itself, etc.), they can charge for verification, which acts as an incentive for issuers to provide more robust credentials and be honest. However, this may compromise the privacy of the SSI system, as issuers may be able to track holders and correlate their activities and interactions with verifiers.

\subsection{Blockchain infrastructures}
\label{section 3.1}

While there are no technical limitations to using other types of infrastructure, the governance and decentralization of trust encourages the use of blockchain in identity management systems. There are different types of blockchains depending on the permissions to join the network as a node (public, private or consortium blockchains) or to interact with it as a user (permissioned or permissionless blockchains), the consensus mechanism, the ability to execute smart contracts, the ability to create private channels or sub-ledgers between nodes, the presence of a native currency/token on the platform, etc.

There are SSI projects, such as \textbf{uPort}, that use public blockchains, e.g. Ethereum, to publish identifiers and DID documents - in the form of smart contract events. Credentials in uPort are JWT (JSON Web Token) issued by one identity to another, the proofs for these claims can also be found as Smart Contract events in the Ether-claim-registry \cite{EthrClaimRegistry}. Ethereum is then used for DIDs, DID documents, and even for publishing attestations of claims\footnote{These are proofs that an issuer has created and signed a claim for a user, not the claims themselves}.
Other SSI projects, such as \textbf{Sovrin}, use permissioned blockchains, which are either a consortium network or a private one. Sovrin uses the Hyperledger Indy blockchain as a DPKI for publishing DID documents, e.g. credential schema and definitions.

Ethereum is used by uPort for storage and for computation as the DID documents are generated from smart contract events, while Hyperledger Indy is used by Sovrin as a decentralized storage component. Blockchains can play different roles in a software architecture \cite{ArchitectureBlockchainApplications}, and depending on what the infrastructure is used for, designers can decide on the most appropriate blockchain platform and configuration that meet the requirements.

Blockchains do not allow backtracking, and data that goes on-chain stays on-chain, it cannot be modified or deleted, and can only be updated by publishing new transactions. This means that the choice of what goes on the blockchain and what does not has to be made carefully, which implies the need for other off-chain components. 
In most SSI systems, personal data contained in credentials are only stored in the user's wallet, but blockchains can be used as a verifiable registry for revocation information, to publish the credential hash as proof of credential integrity and issuance, and for reliable time-stamping, etc.

Public blockchains do not promote access control or other measures to limit access to data. On the contrary, they promote full transparency and openness, meaning that anyone can be part of the network and see the data on the ledger. Transparency and openness are not necessarily bad for privacy, but they make certain use cases where it is necessary to control who can access the data, such as private transactions between two entities, infeasible on public blockchains. It is still possible to conduct private transactions off-blockchain, but this means missing out on the formalized way of conducting transactions on a blockchain and the decentralized trust provided by the platform. This would require a level of data segregation (separating data into silos and controlling access to it on the infrastructure) that not all blockchains or DLTs can provide.
Fortunately, newer generations of blockchain platforms integrate access control and permissions to manage who can be part of the blockchain, who can access what data, and even provide private sub-ledgers maintained by subsets of blockchain members. Hyperledger Fabric, with its membership service providers (MSP) and private channels, is a good example.

Another consideration for public blockchains is that they are more difficult to regulate. From a legislative perspective, it is easier to enforce data protection laws and legal responsibility and accountability on a permissioned blockchain, where nodes are operated by well-defined entities. Public blockchains are generally not suitable for proprietary or government-backed identity management solutions due to the lack of governability, as blockchain nodes can be managed by anyone, anywhere. This is why, for example, various government authorities indicate that permissionless public blockchains are indeed \textbf{hard} to regulate under current jurisdiction, e.g. for financial services \cite{HMT-regulation}.

Blockchains have different levels of decentralization, and the definition of their ecosystem and stakeholders is more operational than technical, but they facilitate working in trustless systems and seem more appropriate for identity management if we want to limit dependency on third parties, which is the case in the SSI model. In terms of auditability, blockchains easily beat any other system. For example, the auditability of blockchains can be done in real time and not retrospectively. An auditor can monitor blockchain transactions in real time and be notified of any events that trigger certain conditions.
In addition, data are redundant between multiple nodes, so blockchains provide fast data recovery and consistent and reliable data across nodes, eliminating the need for reconciliation between counterparties. These counterparties may have different versions of data or truth on a classic web-based infrastructure, but on a blockchain network or DLT infrastructure, all stakeholders running nodes have the same ledger data thanks to consensus. Overall, audits are easier and faster (even in real time) on blockchain networks because the data are trusted and organized into transactions and blocks. Not only do blockchains support classic retrospective snapshot auditing like traditional infrastructures, they also enable ongoing, real-time trusted auditing \cite{deloitte-part3}. Auditability is an important factor for privacy, as it allows regulatory bodies to enforce data protection regulation. 

Beyond the choice of what goes on the blockchain, there is the issue of what - other than identity management - the blockchain platform is used for. For example, the comparison between Ethereum and Hyperledger Indy should go beyond the fact that one is publicly permissionless and one is permissioned. Ethereum also has a cryptocurrency and several other decentralized applications built on top of it, and using these other Ethereum-based applications with Ethereum-based SSI solutions - e.g. uPort - will create a serious correlation problem between user accounts and funds used in different applications on the blockchain platform. Blockchains, which are dedicated only to digital identity management, minimize this correlation risk. The permissioned nature of Hyperledger Indy or Sovrin limits transactions to identity transactions only, and the platform does not require cryptocurrency or fees, as it is generally a consortium between organizations or entities that do not require fees to maintain the network. Although uPort attempts to solve the correlation problem by creating many identifiers and identities (which are Ethereum addresses) for different applications, the need to transfer cryptocurrency to these addresses will cause a correlation between these addresses and the main user account where the funds are held.

In summary, building an identity management system on blockchain requires consideration of the functional and non-functional characteristics of the chosen blockchain platform, as well as the design requirements for building blockchain-based decentralized applications. The need for data privacy and the isolation of data created by different actors requires the use of permissioned blockchains, and keeping personal data off-chain to comply with certain privacy law principles, such as the right to modify or correct data and the right to delete it (for example, GDPR). In its early days, the SSI community considered blockchain as the main infrastructure on which SSI systems should be built. Later, with SSI commercialization and use in different use cases and businesses, SSI matured to include other infrastructures\footnote{Infrastructures may use \textbf{did:web} and \textbf{did:dns} identifiers, DNS-based identifier resolvers, decentralized web registries, etc..} like the web, and most major SSI solution providers have begun to decouple their systems from blockchain dependency in order to be accepted in broader applications.

\subsection{Web-based Infrastructures}
\label{section 3.2}
Web-based infrastructures in SSI use existing Internet infrastructures, protocols, and systems, such as web registries and the DNS protocol, to provide identifiers derived from URLs, create and store DID documents and credential schema on web servers as resources, and resolve them through DNS. These solutions argue that the web is decentralized as an infrastructure, and that each DID owner controls its domain name or servers, which is consistent with the principles of the SSI.

For example, the did:web method uses URLs as identifiers, and DNS as a resolver, and DID documents for each DID owner are published on their own servers as a web resource. While these identifiers may prove useful for sites that can be set up as issuers or verifiers, they cannot be used to identify holders, since holders are not expected to have domain names and manage web servers in order to obtain DIDs. However, this can help bridge the gap between SSI solutions and web sites where issuers and organizations can directly issue verifiable credentials to their users. These credentials are defined by a credential schema and definition found on the server side, along with the DID document and the necessary public keys to verify the signatures on the credentials. Using web-based methods like did:web above or did:dns shows that we can achieve SSI interactions with decentralized identifiers and verifiable credentials, without the need for a blockchain. Furthermore, identifiers like did:key are designed to be extended to public keys or basic DID documents with more direct and efficient resolution, so they do not need an infrastructure in the first place.

Credentials can be defined using URIs, meaning that definitions and schema are treated as web resources, implying the need for a web-based infrastructure and no dependence on a blockchain. Web-based infrastructures offer higher flexibility than blockchains in terms of data management and privacy. For example, it is easier to implement access control to data on the web, and to update or delete data whenever necessary at the request of users ("Right to erasure"\footnote{Article 17. GDPR \url{https://gdpr-info.eu/art-17-gdpr/}.} under GDPR, for example). From a performance point of view, SSI systems that are web-based should perform better because they benefit from the existing technology used to make websites and web-based systems faster compared to blockchains, which typically suffer from low throughput due to decentralization and consensus, etc.

\subsection{Comparing Web-based and Blockchain Infrastructures for SSI}
\label{section 3.3}
The choice of infrastructure and the functionalities assigned to its components depend on the following factors, presented under five (5) big categories:

\begin{itemize}
    \item Satisfy \textbf{the principles of self-sovereignty}: User ownership and control over the infrastructure (by extending these two SSI properties to the infrastructure itself), and decentralization in terms of governance and distribution.
    \item Optimize the \textbf{performance} of the SSI system: Throughput or scalability in terms of how many client operation per second the infrastructure can handle;
    \item Ensure \textbf{data security} on the infrastructure: Availability, data integrity, access control, data segregation, and data confidentiality;
    \item Ensure \textbf{data privacy} for the user: Data updatability, data portability, data deletion, data transparency; 
    \item \textbf{Conformity to legislation} : Auditability and regulability by regulatory authorities in order to ensure legal conformity.
\end{itemize}

These general considerations, which change from one system to another based on requirements, help make decisions about where (blockchain, web-based directory) to put what (DID, public keys, DID documents, resolvers, APIs, and computing elements). Different blockchains also have different characteristics and capabilities to satisfy these factors, so in Table \ref{tab:inf}, we compare blockchains and web-based infrastructures for SSI using the factors described above.

It should be noted that in SSI systems, we rarely find ones that are only blockchain based or only web-based. Typically, we find architectures that combine blockchain with other web-based registries to take advantage of the characteristics of both. Blockchains are perfect as DPKI and as verifiable data registries, but the data that can be put on the chain is limited by legal considerations and by performance and scalability. A key point in the choice of infrastructure is finding a balance between the use of blockchain and non-blockchain components to meet the requirements of a SSI system, both functional and non-functional.

\begin{table}[h]
    \centering
    \begin{tabular}{|*{7}{c|}}
    \hline
         Factor & Public blockchains & Consortium and & Web-based  \\
         &  &  Private blockchains &   \\ \hline
         Infrastructure Ownership & public & private & private \\
         Infrastructure Control & public & authorized nodes & authorized entities \\
         Decentralization & high & low to medium & low to medium \\\hline
         
         Scalability (Throughput) & low & medium to high & high \\\hline
         
         Availability & high & high & medium to high \\
         Data integrity & high & high & medium to high \\
         Access control & low & high & high \\
         Data segregation & low & medium to high & high \\
         Data confidentiality & low & medium to high & high \\\hline
         
         Updatability & low & low & high \\
         Data portability & low & medium & high \\
         Data deletion & no & no & yes \\
         Transparency & high & low to medium & low \\\hline
         
         Auditability & high & high & medium \\
         Regulability & low & high & high \\
         
    \hline
    \end{tabular}
    \caption{Comparing blockchain-based vs web-based infrastructures for SSI.}
    \label{tab:inf}
\end{table}

\textbf{Appendix E} provides more details on the privacy implications of the on-chain and off-chain options for various SSI data elements such as DID documents, cryptographic material, credential proofs, etc.

\section{SSI Identifiers and Cryptographic Material}
\label{section 4}
\begin{figure}[h]
    \centering
    \includegraphics[width=0.8\textwidth]{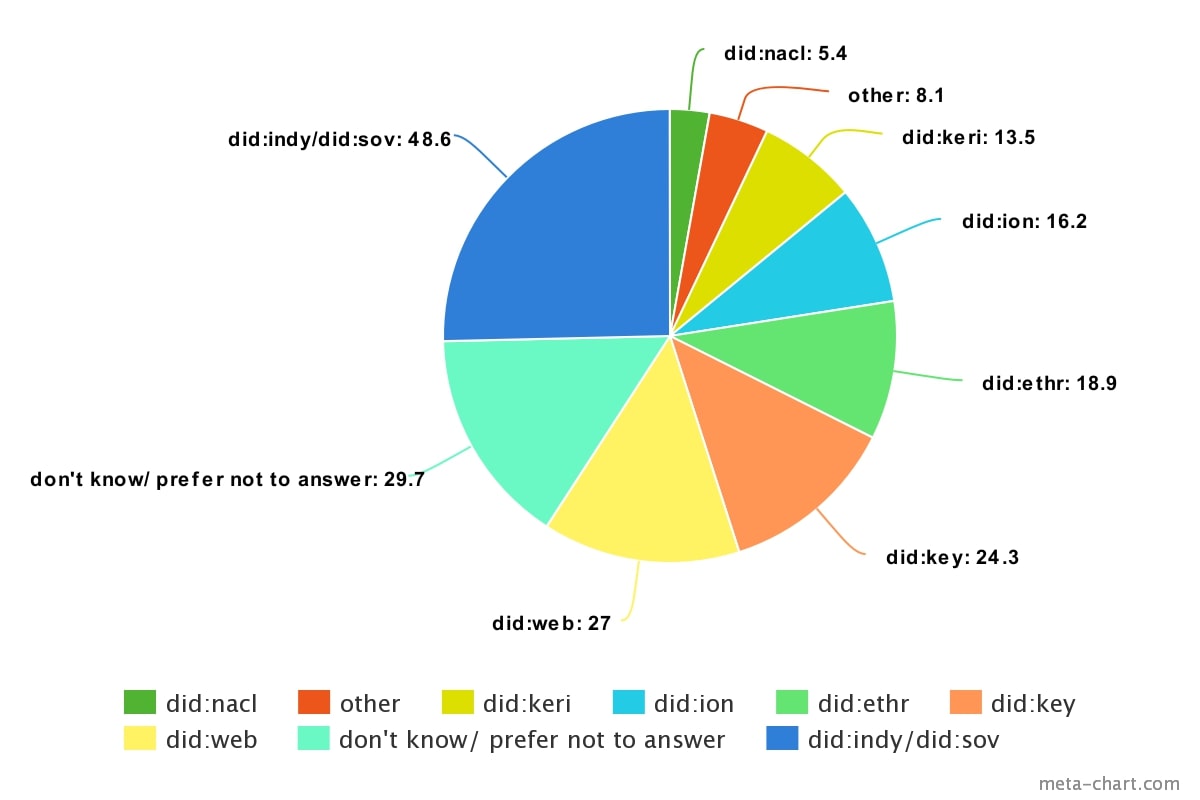}
    \caption{Responses from 37 SSI vendors on DID methods supported by their products [source: CheckD survey \cite{checkD-survey}, reworked.]}
    \label{fig:3}
\end{figure}

The W3C DID standard does not specify a method for generating and resolving identifiers. A DID method specifies how to generate the identifier, how to resolve an identifier to obtain the DID document, and other necessary functions such as delegating an identifier, updating its associated keys, or even deactivating it. There are approximately 100 DID methods currently in use or under development, designed to work with numerous registries and infrastructures \cite{DID-methods}.
A survey carried out by CheckD in 2022 \cite{checkD-survey} shows that the most used DID methods are did:indy and did:sov, followed by did:web, did:key, did:ethr and did:ion, among others, such as did:keri, did:nacl and did:ebsi, which was created by the European Commission for the EBSI and the European Identity Wallet pilot projects. Refer to Figure \ref{fig:3} for a chart of the most used DID methods and \textbf{Appendix B} for a detailed comparison between them. In general, a DID is either for public use or is private.

\subsection{Privacy of DID Methods and DID Documents}
\label{section 4.1}
A DID is always resolved to a DID document. This can be done via a resolver that retrieves or builds a document, either by direct transformation or by expanding the DID itself to obtain a DID document. Different DID methods allow for different content within a DID document, with the most basic only containing a public key, to more sophisticated ones that contain endpoints, multiple public keys, data describing the DID owner, and much more.
Some DID methods only allow the creation of public DID identifiers. For example, the did:ethr method and the did:indy method both rely on blockchain transactions to publish the identifiers to a blockchain ledger. In both cases, the DID documents are built from transactions related to the identifier, like for example transactions that update keys, update endpoints, etc.
A resolver - in this case - needs to get all the transactions and sort them out to create a DID document. However, on a public blockchain like Ethereum, such events are also publicly visible to anyone within the smart contract events \cite{EthereumDIDRegistry}. This means that any did:ethr identifier is visible to everyone, and discoverable. For the different roles in the SSI model, we clearly see that there are identifiers created for public use by verifiers and issuers, and private identifiers intended for holders. A holder can always choose to use a public DID anyway, with additional privacy measures such as: avoiding the use of identity-revealing endpoints in public DID documents, e.g. social media accounts or email addresses, using privacy-preserving endpoints, e.g. shared endpoints, avoiding the use of multiple public keys to avoid correlation, choosing a DID document storage that integrates access control to limit who can access the DID document, etc.

Some DID methods, such as did:key and did:peer, take these considerations into account and provide DID identifiers that are intended for private or peer-to-peer use to avoid the privacy problems associated with public identifiers, but these methods are limited in terms of updatability. These are identifiers that are known only to the entities involved in the communication, and are not published in the ledger or anywhere else. This means that entities must exchange their public keys directly and cannot publish them to any public infrastructure, usually as an out-of-band exchange such as scanning QR codes or exchanging keys via other means of communication such as email or others. Peer DIDs require no infrastructure and are exchanged directly between wallets, they are great for establishing private relationships and secure communication between peers. They are also a good choice for client DIDs, allowing normal users to have more off-chain interactions and more in-wallet identifiers.

Some wallet implementations, such as Talao \cite{Talao}, come provided with a single identifier - following a specific method - and do not allow the creation of more identifiers, or at least the pseudonymization of the unique one. A wallet application may propose different identifiers following different DID methods, or other types of identifiers such as link secrets. A wallet that provides a single identifier limits user privacy because it leads to correlation of user activity and credentials, unless there are methods to hide the DID while presenting the credential.

Blockchain-based DIDs pose privacy risks, as they are discoverable and resolvable by anyone who can read the ledger. This means we can automatically index all identifiers and their associated keys and endpoints and any other data available in DID documents, facilitating correlation and activity tracking. One way to avoid correlation is to allow, if not encourage, users to create multiple DID identifiers for different contexts, relationships, and use cases. Some identifiers, such as did:key and did:peer, require no infrastructure and are perfect for client-side use on their wallets, and do not require a DID document to be stored since it can be cryptographically extended from the identifier itself.

Users can have multiple DIDs, a credential is associated with a single owner through a single DID. This means that when presenting multiple credentials, or claims from different credentials associated with different DIDs belonging to the same user, we risk a correlation problem that defeats the whole need and concept of multiple identifiers.

If all credentials are associated with a single DID, we run the risk of being in a unique identifier case where user activity can be easily traced. If different credentials are associated with different DIDs, then creating a verifiable presentation (cf. Section \ref{section 5.2}) composed of claims from different credentials can prove challenging to do while respecting privacy and not providing a list of identifiers the user owns to a verifier. Some SSI solutions like Evernym's AnonCreds use \textbf{link secrets} to bind credentials to the user to solve the problem of correlating identifiers. However, link secrets do not offer the same capabilities as DIDs, they do not support digital signatures, and they are more proofs of knowledge than proofs of ownership like DIDs with private keys.
Link secrets are a good way to hide the identity of a credential holder within an anonymous credential system, and allow users to solve the correlation problem when combining different attributes or claims from different credentials. When multiple credentials are linked to different versions (blinded commitments) of the same link secret, it is possible to bundle them into a presentation (cf. Section \ref{section 5.2}) that remains verifiable without compromising the DIDs used by the holder. They are more appropriate as a way to anchor credentials to holders than as identifiers themselves. However, link secrets have come under criticism in different papers \cite{9031545}\cite{IdentityWoman} paradoxically because of their unlinkability. In fact, link secrets are known only to the holder, which means that in cases where the holder decides to share them with others, or where the link secret is revealed to another entity that uses it, there is no way for the issuer or verifier, or even the holder, to know that the credentials are being used by other entities. Link secrets are supposed to be used by a single user in all of their credentials for selective disclosure and ZKP, but if two users use the same link secret in their credentials, each of them is able to present different attributes from their credentials and from the credentials of the other without any way to detect it, especially using ZKP. Link secrets are intended to be used in a single wallet by a single user for their credential, but they do not come with any real restrictions on using them in multiple wallets and by multiple users, or in the same wallet by multiple users. Although link secrets are not DID methods, they are widely used in anonymous credential systems implemented in some SSI systems, such as JSON-LD BBS+ credentials and AnonCreds (cf. Section \ref{section 5.1}).

The identifiers and cryptographic material layer - for privacy compliance - must take into account:
\begin{itemize}
    \item different privacy threats stemming from the infrastructure and the DID method. For example, if the infrastructure is a public blockchain with DIDs and DID documents are published as ledger transactions or built from them, curious entities can map and identify all public actors.
    \item the resolver used to retrieve the metadata associated with the identifiers (cf. Section \ref{section 4.2}).
    \item the intended use of the identifier, and the diverse needs of different actors within SSI interactions.
    \item the anchoring of the credentials to identifiers, and the creation of privacy-preserving presentations from such credentials.
\end{itemize}

\subsection{Privacy of DID Resolution}
\label{section 4.2}

In general, depending on the infrastructure used for the DIDs, the resolver is a critical component in the SSI that allows us to resolve identifiers, discover service endpoints, discover communication endpoints, retrieve verification keys to verify signatures, and so on. This resolution is defined in the DID method, and follows the DID resolution algorithm specified in the DID standard. However, there are several ways to resolve identifiers, and each way has its own advantages and disadvantages from a privacy perspective. This means that adopting a resolution method must be considered in light of the inherent privacy aspects of the method, and the additional privacy-preserving measures to be put in place.

For example, the did:indy method resolves a DID by looking for the corresponding NYM transaction on the Hyperledger Indy blockchain. The NYM transaction is JSON data consisting of the DID itself, a verification key (verkey), and optional "diddoccontent", which is a JSON item containing public keys used by the DID owner, DID delegation information, endpoints, etc. The entire DID document can be found in a single transaction metadata.

The did:ethr method, on the other hand, builds a DID document from the smart contract events made by the DID owner. Since the DID is based on the owner's Ethereum address, the resolver looks at contract events (ERC1056 events, for example, for the Ethereum Lightweight Identity) made by the address within a given smart contract, such as the \textbf{EthereumDIDRegistry} contract \cite{EthereumDIDRegistry}. Events consist of invoking functions in the contract, such as delegating control of the identifier, updating keys, endpoints, changing the owner of the identifier, etc. Once the resolver receives events, a DID document is built from these events and returned.

In both cases, the DID document is accessible to any node on the blockchain at first, and to anyone with read permissions to the ledger (everyone for public blockchains). The resolver here can be another smart contract reading the events of another, or a transaction explorer\footnote{For example, Hyperledger Indy TX explorer \url{https://indyscan.io/home/SOVRIN_MAINNET}}. Resolvers for blockchain-based DIDs are implemented either on the blockchain itself, either as external functions that request to read from the nodes. Unlike writing to the ledger or updating DID documents, which is detected and observed by nodes and others via events, reading from the ledger does not require signatures and is sent to only one node, providing better privacy in resolution.

Other DID methods, such as did:web and did:dns, do not rely on public or collectively maintained ledgers. Instead, the resolver relies primarily on DNS. Resolving a did:web consists of creating a URI pointing to a DID document stored as a resource at the web site URL used to create the DID itself. For example, for a DID like \textbf{did:web:\url{www.telecom-sudparis.eu}} the DID document shall be available as a web resource at the address \textbf{https://www.telecom-sudparis.eu/well-known/did.json}. This resolution is based on DNS and inherits all the associated security risks and attack vectors of DNS, so it is recommended to use DNSSEC to avoid manipulation and poisoning of DNS responses. While DNSSEC provides security for correct resolution, it does not provide privacy. 

The DNS server will still track the resolution requests, along with the site itself, since the DID document is a resource on that server. To avoid this, some implementations use the universal resolver to achieve herd privacy or use a VPN. The universal resolver is a resolver set up and operated by the Decentralized Identity Foundation (DIF) and allows resolving multiple DID methods from different SSI solutions such as Indy, Ethereum, Sovrin, Web and others. It can be a perfect API to integrate with an identity wallet to allow users from different SSI solutions to interact with each other.

With these points in mind, enhanced privacy for the identifier layer can be achieved by following these recommendations:
\begin{itemize}
   \item Allow the creation of multiple identifiers that follow different DID methods.
    \item Think ahead to the credential schema, to provide appropriate identifiers for each credential schema.
    \item Implement access control methods on the infrastructure to prevent unauthorized access to DID documents, revocation lists, credential schema and definitions. 
    \item Limit the dependency, if unintended, between the identifier and the infrastructure.
    \item Provide privacy-enhancing techniques for resolving identifiers.
    \item Use privacy-preserving endpoints such as shared endpoints or non-identity revealing endpoints when a holder uses a public DID method.
    \item Provide a private DID resolver. The resolver should be either a trusted entity, an entity that cannot identify and track users, or a service implemented on the user's side.
\end{itemize}

\section{SSI Credentials and Presentations}
\label{section 5}
\begin{figure}[h]
    \centering
    \includegraphics[width=0.6\textwidth]{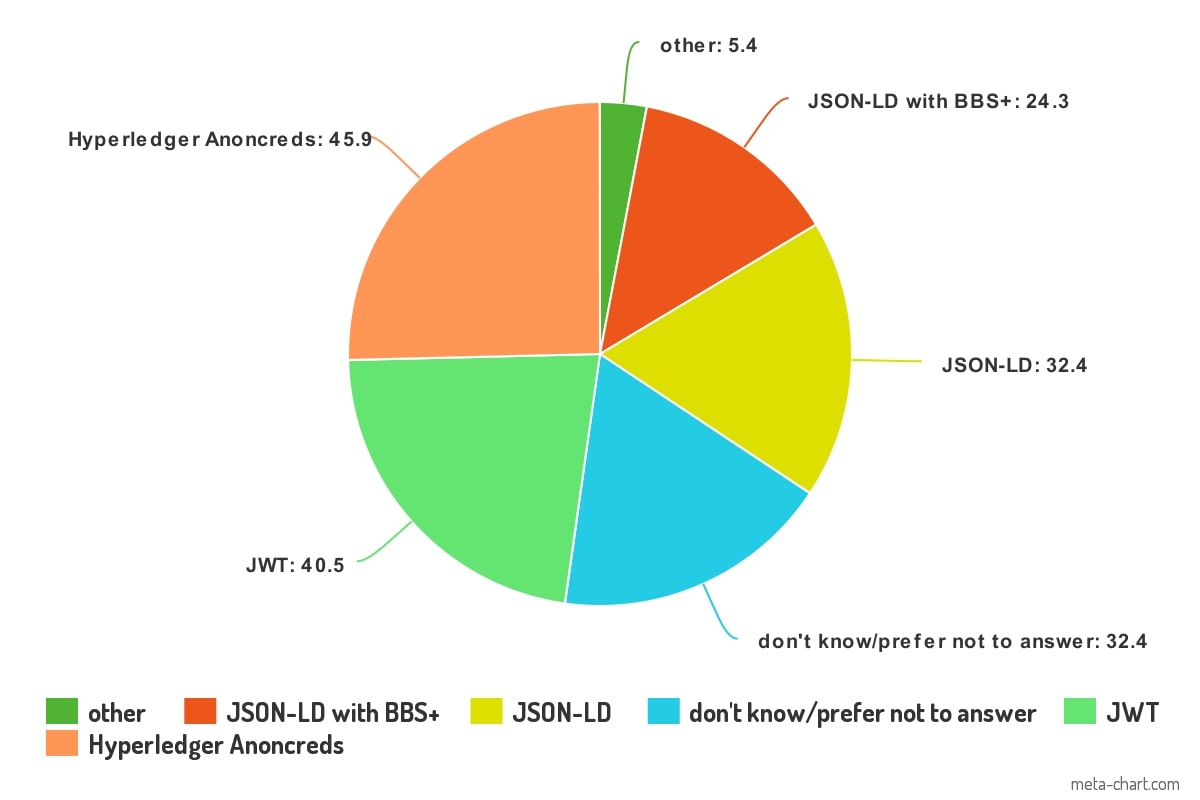}
    \caption{Most used Verifiable Credential schemes [source: CheckD survey \cite{checkD-survey}, reworked.]}
    \label{fig:5}
\end{figure}

While DIDs coupled with an appropriate infrastructure are sufficient to identify and authenticate SSI actors, they are not sufficient for a SSI vision where users can exchange trusted credentials that attest to their attributes, capabilities, qualifications, and other information, either certified by other entities or self-certified. Currently, there are many types of credentials in the SSI that have different flavors, different data representations, and follow different signature schemes\cite{IDwoman}. Figure \ref{fig:5} contains statistics about the most used SSI credentials, according to a survey conducted by CheckD \cite{checkD-survey}.

\subsection{Credentials}
\label{section 5.1}
Credentials generally have two data types: JSON and JSON-LD.
JSON-LD is more complex than JSON because it uses Internationalized Resource Identifiers (IRIs) to link and provide more context to the data. For JSON-LD credentials, the context would be the credential schema and definition, typically stored as resources in web-based registries and referenced in the JSON-LD as URIs.

JSON credentials, on the other hand, must either embed the schema and definition within the credential itself, or include their reference as files or transactions stored on a IPFS or database, or even on the blockchain itself. This is the case with AnonCreds \cite{Anoncreds} and credentials created for Hyperledger Indy, where the credential has a schema published on the blockchain under a given name. Issuers wishing to issue credentials according to the schema will also publish their own credential definition on the ledger under a given name or identifier. Verifiers must take the schema and definition IDs and look them up on the blockchain or VDR (Verifiable Data Registry) themselves to make sense of and verify the credential, unlike JSON-LD where they are just a link away. This creates a dependency between AnonCreds and the Hyperledger Indy blockchain.

In terms of privacy, having the definition and schema on a blockchain means that everyone can see it, even entities that do not need to. For example, including the issuer's public keys in the credential definition and publishing it on a blockchain means that it cannot be changed, and any updates to the definition, such as using new keys, require a new definition to be published. This means that curious entities can track changes and infer that new credentials will be issued following a new definition, and so on. Even with blockchains that support data segregation and private channels (Hyperledger Fabric), while it is possible to have more control over who gets to see the credential schema and definition, we may have possible activity correlation from the issuer side if they are the ones controlling access to the private channels. A workaround for this privacy issue would be to use additional components such as a Permission Manager (or Membership Service Provider in Hyperledger Fabric terminology) that supports ZKP so that the authorized entities can read from the channel without revealing their identities, only their permissions \cite{IDEMIX-MSP-HF}.

JSON LD credentials can put the definition and schema off-chain, which allows more flexibility in terms of credential schema and definition updates, since we only need to change the resource and keep the same URI pointing to it. Off-chain storage also means more privacy in terms of access to the definitions and schema, but comes with security risks. For example, if they are published in a single location such as a server, they may be vulnerable to attack, even though they are generally signed for integrity by the publisher. It is recommended that they be published off-chain, but in a decentralized web-based registry.

It is easier to work with an off-chain definition since blockchains allow little to no flexibility in updating published data. Another consideration is credential revocation, where for credentials like JSON-LD, which allows for simple off-chain components, we can work with web-based revocation lists and registries, whereas JSON based credentials like JWT (JSON Web Token) credentials and AnonCreds use VDR and blockchain ledgers to publish revocation lists and revocation tail files, or use revocation tokens that are signatures to verify revocation. Essentially, JSON-LD credentials are more flexible and dynamic because we can keep the URL and change the resource it points to. On the other hand, JSON credentials are more static. \cite{JSON-LD-Benefits}

Table \ref{tab:vcdata} summarizes the comparison between JSON and JSON LD credentials. We consider the previous comparison points and add some important ones, such as support for multiple languages, signature chaining that allows multiple parties to sign the same credential in a given order (e.g., notarization), reliance on an Internet PKI to resolve URL links, definition of semantics for the credential, number of signature schemes supported, etc.

\begin{table}[h]
    \centering
    \begin{tabular}{|*{3}{c|}}
    \hline
          & JSON credentials & JSON-LD credentials\\ \hline
          Complexity & Simple & Complex \\
          Context support & IANA registry, static or local reference & Any context referred with a URL\\
          Flexibility &  Very limited, contexts are static & Flexible, contexts are dynamic\\
          Graph support & No & Yes (RDF) \\
          Multi-language support &  No & Possible using a language map\\
          Interoperability & Low & High \\
          Support of Linked Data proofs & No & Yes \\
          Signature chaining &  No & Native support \\
          Internet PKI & Not required & Required \\
          Semantics and Disambiguation & Hard and limited & Easy \\
          Signatures supported & Limited & Wide range of signatures \\
          
    \hline
    \end{tabular}
    \caption{Comparing JSON vs. JSON LD Credentials.}
    \label{tab:vcdata}
\end{table}

In the following, we detail two examples from each data format: JWT credentials and Camenisch-Lysanskaya (CL) credentials (specifically IndyCreds/AnonCreds) for JSON, and JSON-LD LDS credentials and JSON-LD BBS+ credentials for JSON-LD. These are the four most used credentials according to the previous statistics. See Appendix C for a more detailed comparison of these four credential types at each stage of their lifecycle.

\subsubsection{JSON JWT Credentials}
\label{section 5.1.1}
JWT credentials follow the JOSE framework \cite{JOSE} signature scheme for proofs and they follow the IANA registry JWT\footnote{IANA JSON Web Token claims \url{https://www.iana.org/assignments/jwt/jwt.xhtml}} for the credential schema and definitions. It is a simple token consisting of a header, a payload containing claims, and a signature. The IANA registry is limited to a hundred attributes, so JWT is very limited semantically, and extending the credential system to more complicated attributes requires statically referencing other registries that contain the credential schema and definition. While this limits the semantics and interoperability of JWT VC, it does make it independent of Internet PKI and DNS resolution of URLs found in JSON-LD. JWT credentials and JSON credentials are generally less dependent on the Internet than JSON-LD credentials, making them suitable for infrastructure-agnostic SSI systems.

JWT Verifiable Credentials provide minimal user privacy, the tokens are base64 encoded, easily decrypted, and contain the issuer's signature to ensure integrity and authenticity, but for the entire credential with no options for selective disclosure or ZKP. A JWT VC contains the holder's DID as part of the claims, so for verifiability we need to reveal the holder's identity (or at least a public key they control) and all the claims, since there are no selective disclosure or zero-knowledge proofs for JWT VCs.

Data minimization in JWT is not achievable via selective disclosure because they do not support the presentation of single claims, since the credential can only be presented as a whole. For example, to present a single claim in JWT VC, the only way to do it is to have a credential that contains a single attribute or predicate that can be shared with verifiers. This means more credentials are needed, rather than a single credential with flexibility to present specific claims. This adds cost to issuers who must sign multiple credentials for a single user, and to users who must store multiple credentials. It is better to use VC schemes that minimize the number of credentials to be signed by issuers and stored by holders, but provide the ability to select attributes to be presented and apply zero-knowledge proofs to attributes as needed.

\subsubsection{JSON Camenisch-Lysanskaya Credentials: AnonCreds as an example}
\label{section 5.1.2}
AnonCreds started as a Hyperledger Indy credential specification called “Anonymous Credentials”. AnonCreds are based on the work of Camenisch and Lysyanskaya \cite{Cam-Lys} which is a formalization of Anonymous Credentials introduced by Chaum in 1982 \cite{AC-Chaum} and the IDEMIX protocol \cite{IDEMIX}. AnonCreds were an attempt by Evernym, a major contributor to open source SSI projects such as Indy, Aries, and Sovrin, to design privacy-preserving credentials for SSI, offering features such as zero-knowledge proofs and selective disclosure. AnonCreds evolved from being part of Hyperledger Indy to a full Hyperledger project on its own, as this credential design is ledger agnostic and can even be implemented without a blockchain. Although in this article we refer to them as "JSON" format credentials, we note that AnonCreds pre-date the W3C VC standard, and that their representation and serialization of credential data are not JSON, but rather their own ZKP CL format, which is represented as JSON when such credential data are later presented by the holder.
As the AnonCreds original developers at Evernym put it, the AnonCreds proofs (presentations) are compliant with the W3C VC standard specifications, while the AnonCreds credentials themselves are not \cite{daniel-hardman}.
AnonCreds are the most widely implemented SSI credentials today. They require a data registry for the credential schema, definition, and revocation tails. The data registry can be a blockchain - like the first implementation on Indy or the Cardano blockchain - and it can also be web-based, like the HTTP AnonCreds method or the did:web method.

The credential schema and definitions in the first iteration of AnonCreds were published on the blockchain, meaning that the credential contains a reference -in the form of a name - to the published definition and schema. The blockchain proved to be more suitable from a privacy perspective for publishing the definition and schema of AnonCreds, since web-based objects, referred to as IRIs in the credential, are only retrievable if the web service is available, and the risk of surveillance by the web service operator is possible.

To hide the holder's identity, AnonCreds issuers bind the credential to a blinded version of a link secret (see section \ref{section 4.1}) known only to the holder. When requesting a credential, the holder includes a cryptographic commitment about the link secret (to prove that they know a link secret without revealing it), and the issuer uses this blinded version of the link secret, along with other elements, such as a nonce, to generate a credential that the holder can present under another blinded version of the same link secret. The credential can be presented more than once to the same verifier without the verifier being able to know the identity of the holder, this is \textbf{\textit{multi-show unlinkability}}.
Another important design choice for AnonCreds are the Camenisch-Lysanskaya (CL) signatures. Attributes in AnonCreds are signed individually and in turn as blocks of messages, a feature supported by CL signatures that solves the problem of creating multiple credentials discussed in JWT VC. Claims remain verifiable even when presented alone or extracted from their original credentials. In addition, predicates are used alongside attributes to achieve ZKP by revealing a predicate about an attribute rather than the attribute itself.

With CL signatures and link secrets, AnonCreds provide better privacy than JWT and other credential types, but they are large, they are computationally expensive to sign and prove, and they rely on the blockchain, so they lack flexibility, portability and updatability. AnonCreds trade performance for privacy, as the privacy they provide is negatively impacted by the low performance of AnonCreds. The overreliance on the blockchain has made projects like Indy, Aries, and AnonCreds unsuitable for large scale identity systems. Independent testers and implementers have reported that it took about 4 hours to issue a batch of 18000 AnonCreds, over 2500 of those failed, and the credential issuance rate (number of credentials issued per second) dropped from 3 credentials per second to less than one credential per second after the first hour \cite{acapy}. This means that the SSI stack Indy-Aries-AnonCreds is not sufficiently scalable in its current form, even though there are efforts like Aries Askar \cite{askar} that can be used to improve the performance of Aries in encrypting, decrypting, signing, and verifying credentials. The same can be said when analyzing the revocation method - using revocation tails published on the blockchain - of AnonCreds (cf. Section \ref{section 5.3.1}).

From a security perspective, a user can have all of his or her credentia1ls stolen if the link secret is compromised. As mentioned above in Section \ref{section 4.1}, link secrets fail miserably at establishing trust between actors if the owner is malicious, because the link secret can be used maliciously to obtain false credentials and attributes, and can be used to impersonate the credential owner without detection \cite{9031545}.
Although AnonCreds are built with privacy by design, they fail to provide a performant credential system for SSI due to an outdated signature scheme, a tight dependency between the credential system and the blockchain infrastructure, and a revocation method that was not designed with scalability in mind.

AnonCreds 2.0\footnote{AnonCreds 2.0 can be found here \url{https://github.com/hyperledger/anoncreds-v2-rs}} as an independent Hyperldeger project may be able to work around the limitations of AnonCreds 1.0 by implementing BBS+ signatures, PS signatures, and changing the way objects and data are described in previous AnonCreds.

\subsubsection{JSON-LD Credentials with Linked Data Proofs}
\label{section 5.1.3}

JSON-LD credentials support multiple Linked Data Proofs \cite{VCDATAINTEGRITY} and Cryptosuites such as BBS+, JOSE and Ed25519. The cryptosuite defines the public key format, the canonicalization mechanism, the hash function used, and the signature formats. These credentials potentially offer a wide range of privacy benefits compared to JWT credentials \cite{JSON-LD-Benefits}, including selective disclosure thanks to ZKP, because they use the JSON-LD data type, which supports these features. However, implemented versions of the JSON-LD Linked Data Signatures (LDS) credentials are incomplete today and do not consider as many privacy features as JSON-LD credentials allows.

\subsubsection{JSON-LD BBS+ Credentials}
\label{section 5.1.4}
Although JSON-LD can support single attribute signing, which can enable selective disclosure, current implementations of JSON-LD with LDS do not support zero-knowledge proofs or selective disclosure. However, using BBS+ as a specific signature brings to JSON-LD credentials the properties of the BBS+ signature, i.e., efficiency in providing better privacy and better performance than most other credential types. There is currently a W3C working draft entitled "BBS Cryptosuite v2023 Securing Verifiable Credentials With Selective Disclosure using BBS Signatures" \cite{W3C-BBS}.

The move to JSON-LD BBS+ credentials is mainly because of the benefits of JSON-LD mentioned above and because of the BBS+ signature scheme. Like the CL signature scheme, BBS+ is a multi-message signature suite that provides key privacy properties such as \cite{DIF-BBS-draft}:

\begin{itemize}
    \item Selective Disclosure: BBS+ allows multiple attributes to be signed while producing a single signature of constant size, making it lighter than CL signatures. The signature is malleable, which means that the holder can generate a proof that shows only the set of attributes they want to show, without disclosing the entire credential.
    \item Zero-Knowledge Proofs: BBS+ supports ZKP. The proofs can even show that a holder has a signature from a particular issuer without showing the signature itself.
    \item Unlinkable proofs: The generated proofs cannot be recognized or linked across sessions with the same or different verifiers. The BBS+ proofs generated from the same signature are indistinguishable. Coupling this with link secrets, we get \textbf{issue-show unlinkability} and a \textbf{multi-show unlinkability} for BBS+ credentials.
\end{itemize}

For canonicalization, the suite uses the RDF dataset normalization algorithm. As for signatures, the BBS+ signatures are compatible with any pairing-friendly elliptic curve \cite{Kumar} such as BN, BLS12, BLS24, KSS16, KSS18, and so on. The W3C draft uses the BLS12-381. The defined suite has two main parts or algorithms: BBS-Signature and BBS-Proof. The draft document from W3C is not yet complete and lacks important sections about the proof part and the privacy and security considerations. However, it is a step towards standardizing BBS+ VC credentials for future adoption by other working groups \cite{BBS}. There are several BBS+ signature scheme specifications, but in general they all cross-reference each other or certain specifications. For example, the MATTR documentation for BBS+ VC references the Identity Foundation draft on BBS signatures \cite{DIF-BBS-draft}.

The key properties of JSON-LD credentials, coupled with how performant and privacy-preserving BBS+ signatures are, make JSON-LD BBS+ Verifiable Credentials look like the next standard in SSI credentials. JSON-LD BBS+ VC breaks ledger dependencies because they are not blockchain dependent like AnonCreds, and because BBS+ is malleable and supports ZKP and proofs of ownership, these credentials do not use predicates that were the core of other ZKP credentials. Allowing the holder to modify the attributes in a way that preserves verifiability but provides more privacy, such as displaying age or information about age from an attribute that is a date of birth, is more practical than creating a predicate about age and signing it. This is more suitable for issuers and holders, allowing smaller credentials with more flexibility and more privacy.

\subsection{Presentations}
\label{section 5.2}
The presentation is derived from one or more credentials held by the user, either by selecting different claims from different credentials, or selecting whole credentials and merging them, or even the simple proof presented by the holder to the verifier to show that he controls or owns or knows a piece of information, e.g. he controls a DID identifier, he owns a credential, he owns a signed attribute from an issuer, etc. The simplest presentation is a VC signed by the controller (to prove ownership) and presented to the verifier, with no changes to the original VC.

When creating a presentation, the user has as much control over what is shared with the verifier as the credential system allows. Some credentials have flexible data structures (e.g., JSON-LD credentials remain verifiable even if the order of the attributes is changed due to the normalization and serialization of the data) and signature schemes (e.g., BBS+ signatures are malleable signatures that remain verifiable after certain changes to the data) where the holder can modify the credential either by deriving predicates, changing the order of the attributes, or hiding certain attributes during the presentation.

There are two important privacy problems that need to be addressed:
\begin{enumerate}
    \item \textbf{P1.} How can it be guaranteed that the verifier does not ask for personal data that it does not need, in order to provide the service it offers?
    \item \textbf{P2.} How can the user create privacy-preserving presentations? 
\end{enumerate}

The first issue is one of governance and liability, where SSI systems need to put in place methods to ensure that verifiers only ask for the necessary personal data and nothing more than what is required to provide services, and that users are informed of what data they need to provide.
The second problem is technical and educational, which can be solved with the right credential system that enables further privacy methods and user education and awareness to master their wallets and use the privacy features available to them.

\subsubsection{Minimize the Data Shared with Verifiers (P1)}
\label{section 5.2.1}
In order to minimize the data to be presented to the verifier and to present only the necessary credentials to access a service, users need to understand what assurances a verifier requires before providing a service. This can be outside the scope of SSI, where users are solely responsible for their private data and who they present it to, or it can be built into the design of SSI itself to ensure that a well-defined context for each transaction between a verifier and a holder is available to the user through their wallet applications.

There are identity management solutions, such as eIDAS 2.0's "European Digital Identity Wallet," that add "trusted lists" of issuers that are certified to provide trust services. These lists help verifiers and users understand who the issuers are and what credentials they provide, thereby increasing trust in the credentials they issue and in the process of obtaining them. Trusted lists are not an original component of SSI systems, however their integration into SSI could prove useful.

The same logic applies to Verifiers. If a verifier is an organization, or acting on behalf of an organization, that verifies credentials presented by a holder to access certain services, then it makes sense to have a public record in the form of a registry that contains information about them. Having a public registry that the wallet agent can reference - before and during the interaction - can better inform the holder about the verifier, what services it provides, and what attributes it needs to provide those services. This is specific to each identity transaction, meaning that each time a user attempts to present certain attributes, the wallet agent can help them select the appropriate attributes and inform them of privacy preserving options.

For example, if a verifier is an adult content provider, or is acting on behalf of one, the public information found about the verifier will help the wallet agent inform the user that it only needs proof of age, not a full credential, and if possible, help the user build a zero-knowledge proof that the user is of age.
This connects the two problems, because even if we have trusted lists and wallet agents that can help users present the right data and maximize their privacy, we still need privacy-enabling methods implemented on the wallet and on the credentials.

\subsubsection{Privacy-Preserving Verifiable Presentations (P2)}
\label{section 5.2.2}

In a presentation, even if the identity of the holder is not revealed, they must be able to prove their ownership of the attributes presented. This is done by using digital signatures generated by the private key associated with the holder's DID mentioned in the original credential, or by proving knowledge of a link secret present in the original credential. The first case implies revealing that the holder is has multiple DID identifiers if the attributes are taken from different credentials that are anchored to these DIDs. This implies a correlation between the user's identifiers and undermines the concept of multiple compartmentalized identifiers that SSI promotes. In the second case, credentials that share the same link secret, but in different blinded versions, are safer to combine into a single presentation because they do not reveal more information about the user or create a correlation problem (cf. Section \ref{section 4.1}).

Privacy is a big issue for verifiable presentations. In the W3C standard, the section for privacy \cite{VCPrivacy} covers about 18 points where privacy can be undermined.

\subsection{Revocation}
\label{section 5.3}
Revoking is an important part of any credentialing system. Credentials may have a limited life span, or they may simply be revoked upon renewal, compromise, issuance of false credentials, or non-respect of credential usage. Revocation can be temporary (suspended) or permanent (revoked).

When verifying the presentation of a credential, the verifier must also be sure that the credential is still valid and has not been revoked. This makes non-revocation itself a claim to be verified, and there are many ways to prove such a claim. For example, web certificates are revoked by their issuer, the Certificate Authority, via Certificate Revocation Lists (CRL) containing lists of revoked certificates, or via Online Certificate Status Protocols (OCSP).

Credentials can be compared to certificates, with some differences, notably in the authority model (often decentralized and not based on an authority) and in the way revocation is implemented. For example, in many SSI credential schemes, the proof of non-revocation is computed by the holder, using revocation information transmitted by the issuer during credential issuance, and using revocation data available in the ledger. This means that the holder has enough autonomy to build their own non-revocation proofs without relying on a third party between them and the verifier (unlike, for example, OCSP).

\subsubsection{Revocation Methods (Part of the Credential Scheme)}
\label{section 5.3.1}
As the analogy with X.509 shows, VC has several methods to implement the revocation aspect of a credential, although not all credential schemes take revocation into account.

For example, the revocation policy of an "AnonCreds" credential is specified in the credential definition\footnote{\url{https://hyperledger.github.io/anoncreds-spec/\#generating-a-credential-definition-with-revocation-support}} and generally uses revocation tail files, with a cryptographic accumulator that is periodically updated by the issuer and published to the ledger. A revocation tail file contains multiple randomly generated cryptographic factors, indexed by line. Each factor is a large number whose index is assigned to a credential issued by the issuer, and when the credential is revoked, the current accumulator value is multiplied by the factor associated with the credential. If the credential's factor "contributes" to the accumulator's value, it means that the credential is not revoked. Revoking a credential requires the issuer to update the accumulator, essentially removing factors that no longer contribute and adding new factors as new credentials are issued. When they receive the credential, only the issuer and the holder know the index in the revocation tail file that corresponds to the credential. The revocation tail file and the value of the accumulator are public and published by the issuer. A verifier does not know which factor corresponds to which credential, and only the issuer or holder is able to generate a proof of non-revocation. 

Revocation tails must be large enough to provide privacy by obfuscation. However, large revocation tails have a negative impact on performance. Measurements\footnote{Measurements can be found here \url{https://github.com/bcgov/indy-tails-server/blob/main/README.md}} show that the larger the revocation tails, the more time it takes to generate non-revocation proofs. Metrics show that for a revocation tails file of size 2.6 Mo, containing only 10000 revocation entries, it takes about 5 seconds to generate a proof. Aries Cloud Agent (used to issue, verify and exchange AnonCreds) sets the revocation tails limit to only 32768 entries. 

AnonCreds 2.0 \cite{Anoncreds2.0} replaces revocation tails with a Merkle tree that is either published on the ledger or off, depending on the infrastructure. Credentials are indexed, where indices of non-revoked credentials are the leaves of the Merkle tree. In addition, an entirely new revocation element called "ALLOSAUR" \cite{cryptoeprint:2022/1362} is added to the credential scheme. This revocation is highly scalable since it has a constant time to verify and a constant size for elements. The proof of non-revocation is a proof of membership showing that the credential belongs to non-revoked credentials (if it is an inclusive revocation), and a "false" membership proof is computationally infeasible, meaning that this revocation method is better in terms of performance and security than revocation tails.

ALLOSAUR even supports ZKP of non-revocation, where even the verifier can check a value without revealing which value he is checking, making it better from a privacy perspective. However, it requires the introduction of a new role/component in the ecosystem called "accumulator manager" in ALLOSAUR that interacts with issuers and holders to update the revocation registry, and the witnesses used by the holder to prove that their credential has not been revoked \cite{cryptoeprint:2022/1362}. This revocation manager allows issuers to create lightweight revocation registries and scale them to handle millions of credentials, solving the performance problems of AnonCreds 1.0. Privacy is better, assuming the revocation manager is secure and trustworthy, but it introduces another component that can be somewhat centralized, as it is a web service (service with HTTP interfaces) that introduces off-blockchain steps for credential verification, unlike the original AnonCreds, which is blockchain-based. However, this method is more suitable for the direction that AnonCreds and other SSI credentials are converging towards, which are JSON-LD based credentials with more mature signature schemes such as BBS+ signatures.

The BBS+ credentials proposed by MATTR seems to avoid all this complication and engineering of revocation methods based on cryptographic proofs and accumulators, and opts for a simple, transparent, interoperable method, being revocation lists\footnote{\url{https://learn.mattr.global/docs/profiles/web/revocation}}. In matter of fact, JSON-LD BBS+ credentials simply use the Revocation List 2020\footnote{\url{https://w3c-ccg.github.io/vc-status-rl-2020/}} of W3C, where the revocation status of credentials is published. The privacy of these revocation lists is herd privacy, which can only be achieved when there is a large number of issued credentials. The minimum revocation bitstring is at 16KB uncompressed. Verifiers can increase privacy for holders by caching the revocation lists, i.e. the verifier's access pattern will not be inferred to the issuer. From their side, issuers can also improve privacy by using content distribution networks (CDNs) to make the revocation lists available at the end-device level, reducing the need to request lists from issuers.

JWT credentials use revocation lists to mark the credentials as revoked, or wait for the expiration date specified in the credential.

\subsubsection{Revocation Proofs (Part of the Presentation Proofs)}
\label{section 5.3.2}
As the previous subsection shows, different revocation methods implemented at the credential level have different implications at the presentation level. For example, revocation methods handle these two considerations differently: (1) \textit{who is responsible for generating non-revocation proof?} and (2) \textit{how much the issuer is involved in non-revocation verification?}.

In AnonCreds 1.0 and 2.0, the \textbf{holder} is the entity that generates the non-revocation proofs based on the value of the accumulator and the cryptographic factor associated with its credential, known only to it and the issuer. If access to the tail file or the accumulator is public, the issuer is not involved in non-revocation verification. If access to the tail file or the current accumulator value is controlled by the issuer, they may be able to correlate verifier activity, but never holder activity. However, the revocation manager introduced in AnonCreds 2.0 adds another entity besides the issuer and holder that knows certain values such as the witness, accumulator, credential factor in the revocation tail file, etc, but this entity has no access to or knowledge of the credential itself.

The revocation lists used in JWT credentials and in JSON-LD BBS+ and others are pretty straightforward: revocation is checked by the \textbf{verifier} directly against a list provided by the issuer. The holder has little to no involvement in this non-revocation verification, so it is up to both issuers and verifiers to take safeguards to protect their privacy. As seen above, holders have herd privacy if the revocation list has many credentials listed, verifiers can use local caching of lists to improve the privacy of their verification, and issuers can use CDNs to avoid being solicited every time a verification is in progress. Overall, revocation lists avoid complex cryptographic computations required by the holder in the case of revocation tails or Merkle trees, but they imply less sovereignty for users and direct revocation verification between issuers and verifiers, putting the holder aside.

Revocation must be considered as an extension of the credential itself, it is one more claim that the credential is still valid at the moment of presentation. This means that the revocation method should be in synergy with the credential's scheme, for example, if the credential's scheme is an anonymous credential scheme (like AnonCreds or IDEMIX), the revocation method should also be designed to preserve the credential holder's privacy without revealing additional details.

\section{SSI Wallet Applications}
\label{section 6}

Traditional physical wallets are made by a manufacturer, sold in a store, purchased by a user, and used to store physical credentials. These are identity documents that are used by their owners to identify and authenticate themselves to various service providers. The notions of ownership and self-custody for physical wallets are easy to think about, as they are truly owned by and under the full responsibility of their users. 
However, their digital analogues are not so easy to think about in terms of ownership. First of all, there are many entities involved in the development of these applications, the components and the chosen architecture, the devices on which they are installed, the communication protocols they use, the identifiers and the keys they generate, and so on.

\begin{figure}[!htb]
\begin{minipage}{0.48\textwidth}
    \centering
    \includegraphics[width=1\textwidth]{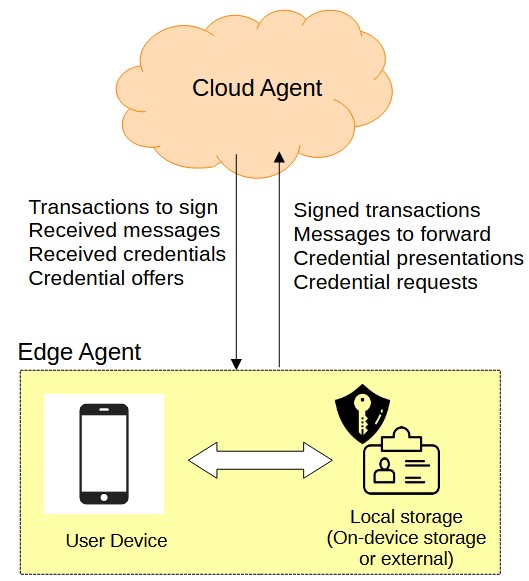}
    \caption{A cold wallet with an online component responsible for routing transactions and communications between the user's wallet and the rest of the SSI components.}
    \label{fig:6}
    \end{minipage}\hfill
   \begin{minipage}{0.48\textwidth}
    \centering
    \includegraphics[width=0.93\textwidth]{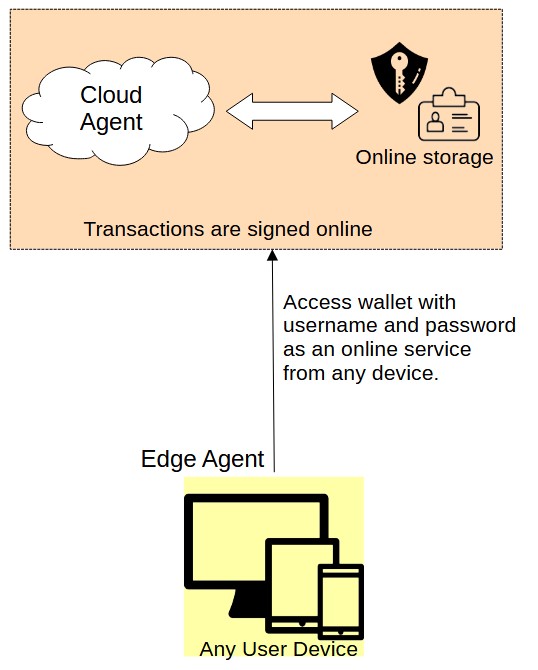}
    \caption{A hot wallet, provided as an online service by an online wallet provider that is responsible for generating, storing and signing the user's transactions on the user's behalf when requested.}
    \label{fig:7}
    \end{minipage}
\end{figure}

We distinguish two main types of digital wallets: \textbf{hot wallets}, which are online wallet services hosted in the cloud, and \textbf{cold wallets}, which are installed on the user's device with local storage for keys and credentials. SSI wallets are generally cold wallets, as the applications come with a local component called a \textbf{edge agent}, which is installed on the user's device, and an online component, called a \textbf{cloud agent} or mediator, that acts as a gateway between the wallet and the SSI infrastructure and as a mediator between different edge agents. Separating the cloud agent from the edge agent is an important logical procedure, as it allows us to better understand the privacy of a wallet application (cf. Figures \ref{fig:6} and \ref{fig:7}).

\textit{PS: Some, like the cryptocurrency wallets communities, consider any internet-connected device to be a hot wallet. In this paper, we think of digital wallets as a spectrum that goes from hot - fully online - to cold - fully offline.} \\

\subsection{Hot Wallets and Cold Wallets}
\label{section 6.1}
Hot wallets are convenient in terms of performance and ease of use because they are hosted on online servers that offer more computing and storage capacity than a user's personal devices. They provide users with easy access to their credentials, keys, and identifiers from any device, as well as easier backup and recovery processes. They also provide a smooth exchange of credentials and messages between users. On the downside, hot wallets violate SSI principles such as ownership and self-custody over credentials and identifiers. Generating, storing, and managing credentials all on a remote cloud service can prove to be a far cry from the concept of self-sovereignty, as all SSI functionality relies on the cloud wallet provider, which takes control and acts on behalf of the user.

Cold wallets take the generation, storage, and management of keys and identifiers offline by implementing the required functionality in a mobile or desktop application (sometimes even a browser extension) and relying only on an online agent - \textit{if required} - to publish identifiers, perform updates to DIDs and DID documents, or exchange credentials and messages between users. This means that - as shown in Figure \ref{fig:6} - a cold wallet will have a computational capability to generate keys and identifiers and create digital signatures, and local storage, which may be the memory of the device on which the application is installed, or a separate storage component such as a SIM card or SD card.

Depending on where we implement functionality like generating and storing the cryptographic keys, and where we store credentials, etc., a wallet can range from a fully hot wallet where all functionality is done online, to a fully offline hardware wallet with little to no online functionality. For example, Hardware wallets like \textbf{Trezor wallet} \cite{Trezor} or \textbf{Ledger Nano} \cite{Ledger} are as simple as USB sticks with extra firmware to generate and store keys, with some connectivity such as Bluetooth or USB cable to communicate with a computer or a mobile phone, and to interact with other wallets and the infrastructure. Other wallet solutions provide specific hardware or software - like \textbf{Talao} \cite{Talao} and \textbf{Dock wallet} \cite{Dock} - or sometimes both, like \textbf{Foundation}'s \textit{Passport Hardware} and \textit{Envoy mobile application} \cite{Foundation}.

\begin{figure}[h]
    \centering
    \includegraphics[width=0.7\textwidth]{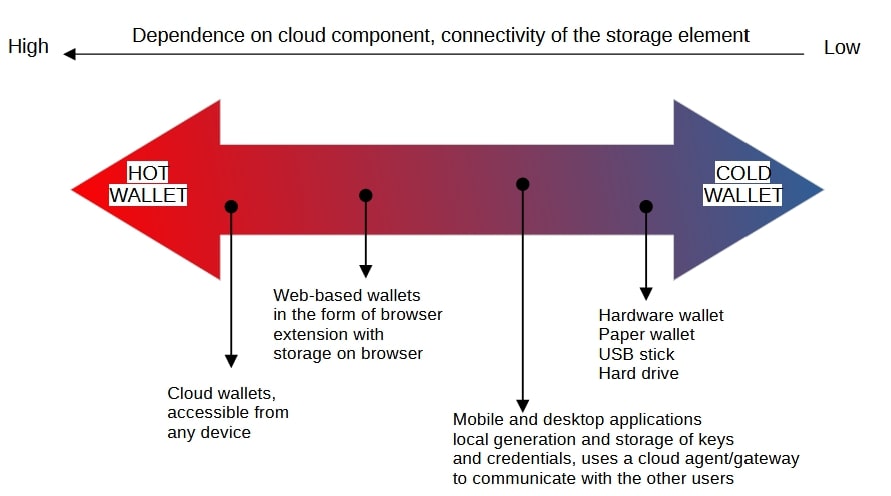}
    \caption{Spectral representation of wallet applications, from fully cold wallets to fully hot wallets.}
    \label{fig:8}
\end{figure}

In the middle of the spectrum shown in Figure \ref{fig:8}, we find wallets with an online component and an offline component, e.g.. \textbf{Metamask wallet} \cite{Metamask} which runs as a browser-extension (more about it below). The left end includes wallets that are fully online, hosted by a cloud service, and accessible to users via a username and login. In such cases, all creation and storage and communication functions are performed online. Between Metamask and fully hot wallets with no local storage, we find several examples like \textbf{Exodus wallet} \cite{Exodus}, \textbf{Coinbase wallet} \cite{Coinbase} and others. To correctly position a wallet, we need to identify which part of the wallet application (edge/client side - cloud/server side) does what functionality.

\subsection{Cloud Agents}
\label{section 6.2}
The cloud agent is the component of the wallet that allows the user to interact remotely with other entities, such as issuers and verifiers and other holders. Apart from cases where the entities can physically interact by scanning QR codes and directly exchanging credentials via NFC or Bluetooth or any kind of local connectivity between devices, a cloud agent is required to exchange credentials and eventually messages between users. A cloud agent has several functionalities such as implementing resolver APIs to handle DID resolution requests coming from edge agents, providing APIs to read and write to the blockchain such as publishing new DIDs or updating published keys and DID documents, forwarding messages and credentials between edge agents, storing messages when the edge agent is unavailable, etc.

Many cloud agents SDK and tools are open source, e.g. Hyperledger Aries \cite{Aries} and Veramo \cite{Veramo}, and can interface with various storage and key management systems. Other existing wallet solutions, e.g., Metamask, have a \textbf{edge agent} that holds the user's keys locally (e.g. on the browser) and a \textbf{cloud agent} - e.g. Infura \cite{Infura} - to communicate with the blockchain network of choice and cloud storage platforms like IPFS.

The dependency on the cloud agent is the factor we use to place a wallet solution on the hot/cold spectrum in Figure \ref{fig:8}. There are many privacy considerations that must be taken into account to determine how much a cloud agent should be involved and depend on, but also performance considerations that sometimes make the need for a cloud agent unavoidable. For example, wallets that are designed with messaging capabilities and remote proofing will need a cloud component to handle these online transactions. Other wallets, such as those designed for in-person credential exchange via QR codes, do not require a cloud agent in most cases. However, what is generally required for wallets is the ability to resolve DIDs to obtain public keys, proofing methods and communication endpoints, and this requires a reliable resolver that must be able to resolve - in several cases - different DID methods. This means Internet connectivity and read access to various blockchains and infrastructures.

Cloud agents must implement privacy and security methods that protect transactions coming from the edge wallet component. For example, some outlines \cite{ARF} require that the wallet provider - which here is generally the application developer responsible for both the edge agent and the cloud agent - provide privacy guarantees such as not tracking user activity of credential exchange, not colluding with other entities to correlate user activity, and not being able to collect information about the use of the wallet that is not necessary to provide the wallet services, along with keeping a record of all data processed by the wallet provider about the user. The wallet holder also has the right to access the data the Wallet Provider holds about them and to restrict and object to the processing of such data, and any data held about them should be deleted if they deactivate their wallet.

\subsection{Edge Agents}
\label{section 6.3}
The edge agent is a term used to describe the parts of the wallet application that are hosted on the user's personal devices. This can be a mobile application, a desktop application, a browser extension or a third party device (hardware, software) that can interface with a computer or mobile phone to communicate with the online agent. An edge agent is typically built to store and compute - primarily key generation - in a secure environment that is isolated and protected from other applications and from the Internet. Depending on the implementation, the edge agent may be used to generate and store DIDs and their associated keys, sign credentials or presentations, and store or generate credentials. It is also used to generate connections (pairwise and peer DIDs) and store connections between peers.

The privacy of the edge agent and the security of its computing and storage capabilities are heavily dependent on how well the application is isolated on the user's device. It is common for edge agents to use HSM (Hardware Security Module), SE (Secure Environment), TEE (Trusted Execution Environment) and rely on more isolated components such as SIM cards to generate and store sensitive information such as private keys or to sign credentials. In addition to using these security features, edge agents must incorporate privacy and confidentiality measures to protect the data they store from other applications installed on the same device.

Edge agents may also overlap with some of the functionality typically associated with cloud agents. The edge agent is installed on a device with Internet connectivity, the intermediary cloud agent can even be removed in some peer-to-peer architectures where peers can directly read and write to the infrastructure - blockchain or web-based - and handle wallet to wallet communications without going through a message relay or a cloud agent. If the connectivity and networking part is handled by the edge agent, this implies further security and privacy considerations for the design of the edge agent. Following the same logical separation as in the hot wallets and cold wallets subsection, we can position a wallet in the spectrum between the two based on functionalities attributed to both the cloud and the edge agent.

In Appendix D, we enumerate critical and essential functionalities common to most SSI wallets and analyze their attribution to either the edge agent or the cloud agent.
 
\subsection{Wallet to Wallet Communication Protocols}
\label{section 6.4}
SSI wallets can communicate using traditional protocol stacks based on HTTP/HTTPS such as RESTful API architectures, or they can communicate using P2P protocols such as IRC. The SSI community has defined the DID-comm and DID-Auth protocols to handle communication and exchange between two DID owners or more, these are transport-agnostic protocols, meaning that they are suitable for both peer-to-peer and server-based contexts, capable of working over HTTP/HTTPS, web-sockets, IRC, Bluetooth, NFC, etc.
We focus in this section on protocols designed specifically for SSI: DID-Comm and DID-Auth.

\subsubsection{DID-Comm Protocol}
\label{section 6.4.1}
DID-Comm started in the Hyperledger Aries RFCs \cite{ARIES-RFC} which describes a way to exchange messages between two DID identifiers. The second version, DID-Comm 2.0, is a protocol by the Decentralized Identity Foundation (DIF) \cite{DIDComm}. The two versions are very similar, but have some differences in the headers and media types defined. A comparison between the two can be found in \cite{V1VSV2}.
DID-Comm enables the exchange of messages and credentials between two DID owners, in a secure manner based on the DIDs, their associated cryptographic material and the DID documents.

\begin{figure}[h]
    \centering
    \includegraphics[width=0.8\textwidth]{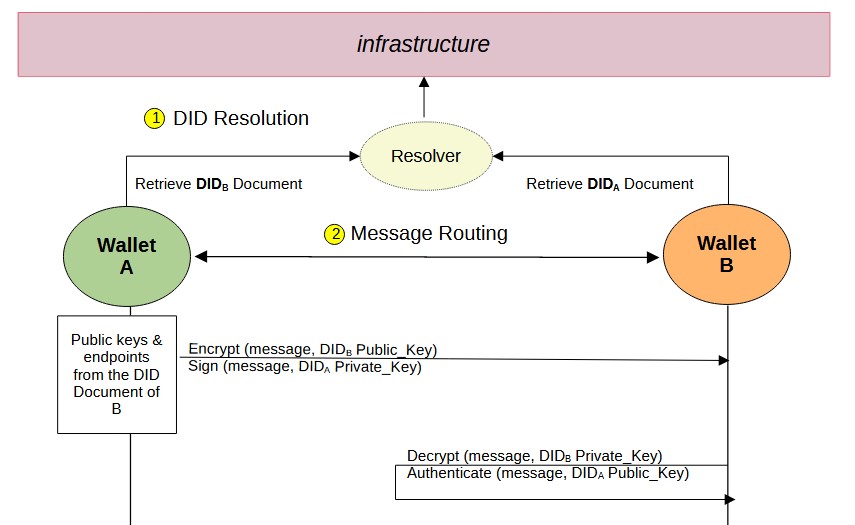}
    \caption{Overview of DID-Comm flow, with highlighted security and privacy crucial points for the protocol.}
    \label{fig:10}
\end{figure}

Figure \ref{fig:10} gives an overview of the protocol. It shows two wallets A and B communicating using two DIDs DID-A and DID-B. Each wallet must resolve the DID of the other wallet (step 1) in order to get the public keys and endpoints that will be used to encrypt, authenticate and exchange messages between the two wallets.
The protocol uses encryption envelopes to secure messages using the keys of the DID owners, with different types of encryption. These envelopes are essential for protecting messages sent between edge agents, often through cloud agents and possibly messaging servers. The fact that the encryption itself is based on the DIDs makes the SSI system reliable and self-sufficient in terms of privacy, since it does not rely on external privacy and security services, such as e.g. HTTPS or TLS, but on existing technologies and protocols that come with the SSI standards. This makes it easier to set up SSI systems that protect confidentiality and authenticity of communication without relying on external protocols.
The second half of the DID-comm protocol (step 2) handles message routing between the two wallets. The DID-Comm protocol is transport agnostic, meaning that it can use HTTPS Post or libp2p\footnote{The peer to peer network stack, \url{https://libp2p.io/}
} for example. In addition to transport, DID-Comm also defines a routing protocol in which messages are forwarded between cloud agents (mediators) that are partly trusted by the sender and the recipient (\cite{DIDComm}, Section 9.4 \#Routing Protocol 2.0). Messages are encrypted using envelopes, so mediators do not know the content of the message.
The privacy of DID-Comm depends on:
\begin{itemize}
    \item The privacy of the DID resolution (step 1 in Figure \ref{fig:10}) between the two DID owners, this includes the privacy of the endpoint to which the message should be delivered, the ability in some DID methods to track who resolved which DID (as in DID-web), etc.
    \item The transport layer (step 2 - message routing - of Figure \ref{fig:10}) between the wallets exchanging messages, including how agents route messages from the sender to the receiver, the possibility of intercepting and decrypting messages, the privacy of routing, such as whether agents may know the addresses of senders and receivers, etc.

\end{itemize}

Back to the message envelopes in DID-Comm, in both V1 and V2 of the protocol there are different types of messages: \textbf{signed-encrypted, anonymous-encrypted, authenticated-encrypted, and plaintext messages}.
Plaintext messages have no encryption. The other message types are based on encrypting the message or putting it in an encrypted envelope using the public key associated with the recipient's DID. If that public key is published or known to many entities, we can have anonymously encrypted messages that only the recipient can decrypt, but there is no way to tell who the sender is. If the sender additionally signs the message with his private key, the message is signed-encrypted, and if the messages are exchanged in an authenticated channel between the two entities, it is an authenticated-encrypted message. The sender is known in both signed-encrypted and authenticated-encrypted. However, although signature proves integrity and authenticity just as authentication does, we use signed-encrypted messages for cases where the origin of the message needs to be proved to a third party or the message indicates some sort of transaction between the two entities (signature proves intent and consent).

Another different envelope is the signed unencrypted envelope, which contains a non-repudiable signature that authenticates the origin of the message and is generally used for public challenges (such as scanning a QR to request a credential) or messages intended for multiple recipients, even if some or most of them are unknown.

These different messages enable different interaction capabilities between entities. For example, anonymous encrypted messages outside an authenticated channel allow anonymous message exchange, and encrypted messages inside a channel authenticated with two peer DIDs created for a one-time exchange or between two entities allow pseudonymous message exchange.

\subsubsection{DID-Auth Protocol}
\label{section 6.4.2}
DID-Auth allows two DID owners to authenticate themselves using their DID identifiers, either one way or both ways, as it supports mutual authentication and one way authentication. This helps to create a secure mutually authenticated channel. It is a challenge-response protocol that builds a challenge starting from the DID document that contains endpoints and "authentication" filed that specifies how the DID owner wants and can be authenticated. The owner is challenged to prove that they control the DID, the challenge being constructed in such a way that only the controller of the private key associated with the DID is able to solve it, thereby proving that they do indeed control the DID they present. The challenge can be presented as a link, QR code, or any other form. DID-Auth, like DID-comm, is transport agnostic and can be delivered to any endpoint specified by the DID controller or owner in their DID document.

The first step in authentication requires resolving the DID, retrieving the required endpoints and authentication information, building the DID-Auth challenge, and delivering it to the specified endpoint.
The challenge may also include proof of control over the challenging party's DID and must include a cryptographic nonce that makes the challenge unique and prevents replay attacks and eavesdropping. Again, the security of this protocol relies on the security of the DID resolution.

A wallet using DID-Auth can be configured so that the user agent allows the user to choose which DID to use for authentication, and it can also be configured to generate new DIDs for each new connection to another (pairwise DIDs) entity. DID-Auth allows for re-authentication and step-up authentication between two entities that have already mutually authenticated, where one party can ask the other to re-authenticate or present a public DID or credential to step-up their authentication level (or level of assurance in their identity). This is a common practice for financial and identity-related transactions, where the initial authentication generally allows only a limited number of transactions, and the performance of additional transactions - such as applying for a credential such as a passport - requires the user to step-up by presenting other credentials or re-authenticating with a different DID.

Now that we have provided a deep review of SSI ecosystem and components, we move to introduce DAD, our Design Assistance Dashboard that provides a fast reliable way to analyze and evaluate SSI systems in terms of privacy.

\section{Design Assistance Dashboard (DAD) Framework}
\label{section 7}
DAD is a practical application of our layered model, providing a review tool that classifies the privacy of different SSI components and their interdependencies. In this section, we present our DAD and how to apply it.

\subsection{DAD Definition}
\begin{figure}[h]
    \centering
    \includegraphics[width=1\textwidth]{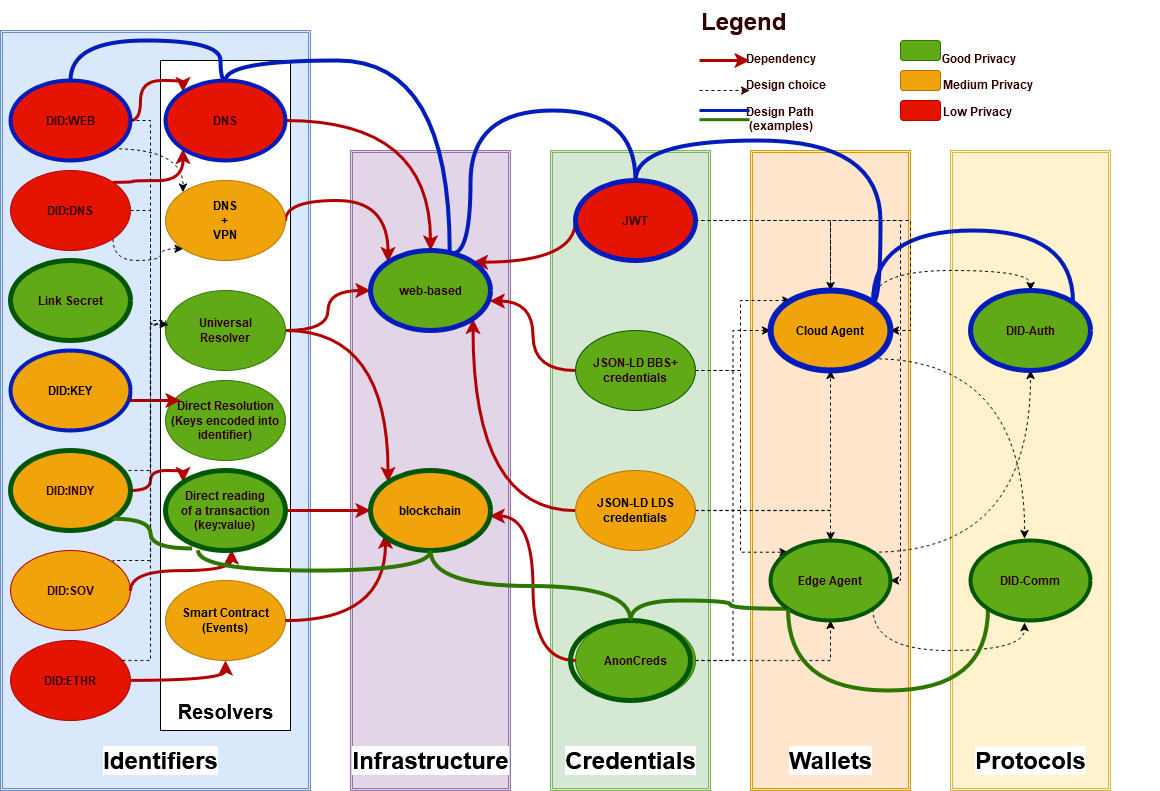}
    \caption{SSI design assistance dashboard with two example design paths (in blue and green). }
    \label{fig:11}
\end{figure}

The privacy-oriented layered framework can be represented as a design dashboard if different components in each layer are enumerated and associated with other components from other layers. This association either represents a dependency or a possible design choice, so by systematically associated SSI components across layers to each other, we end up with a dashboard that graphically illustrates SSI design process.
Figure \ref{fig:11} shows our dashboard display, with two example design paths, marked in blue and green. 
Rather than providing an exhaustive representation of dependencies and design paths, the intention is to offer a visual representation of SSI design dependencies and paths, as well as a colour scheme that allows the level of confidentiality associated with each component to be identified at a glance.

The dashboard is used by selecting different components from each layer and linking them together to build an SSI system. Dependencies are not optional links, i.e. if a component is selected from one layer and has a dependency on a component from another layer, then both must be selected. Note that multiple components of the same layer can be selected in an SSI system. For example, a system may have multiple types of identifiers, use different types of infrastructure, and offer different types of credentials.

\mycomment{Most SSI identifiers have a functional dependency on either blockchain, DLT, or web-based infrastructures. Some DID methods and identifiers do not require infrastructure, such as did:key and link secrets. Credentials also depend on the identifiers used to bind credentials to their issuers, their holders and the associated keys used to sign such credentials. They also depend on the infrastructure used to publish the credential schema and definitions, or even to publish credential proofs or hashes. Finally they depend on the wallet and its protocols for storage and exchange. The wallet relies on the communication protocols, both between wallets (wallet to wallet) and between wallet components themselves (edge to cloud).}

To introduce a privacy indicator in DAD, we refer to our earlier analysis of different components in relation to each other. In line with the privacy requirements a component should provide in a layer, we identify three (3) levels of privacy ranging from low, medium, to good. This privacy level only indicates the component's ability to provide the privacy features that have been mapped to the layer in which the component operates. Note that certain components - such as did:web - may present a low level of privacy as an identifier when used for a normal user-holder, but are still relevant because they are intended to be used for public actors, such as issuers and verifiers, and do not need to meet the privacy requirements mapped to their layer.

\subsection{Applying DAD: A step-by-step guide}
Path 1 and Path 2 (in blue and green) illustrated in Figure \ref{fig:11} are the result of applying the step-by-step guide to two use cases presented below. They should not be taken as recommendations and do not represent the best design choices for privacy. 

\begin{figure}[h]
    \centering
    \includegraphics[width=1\linewidth]{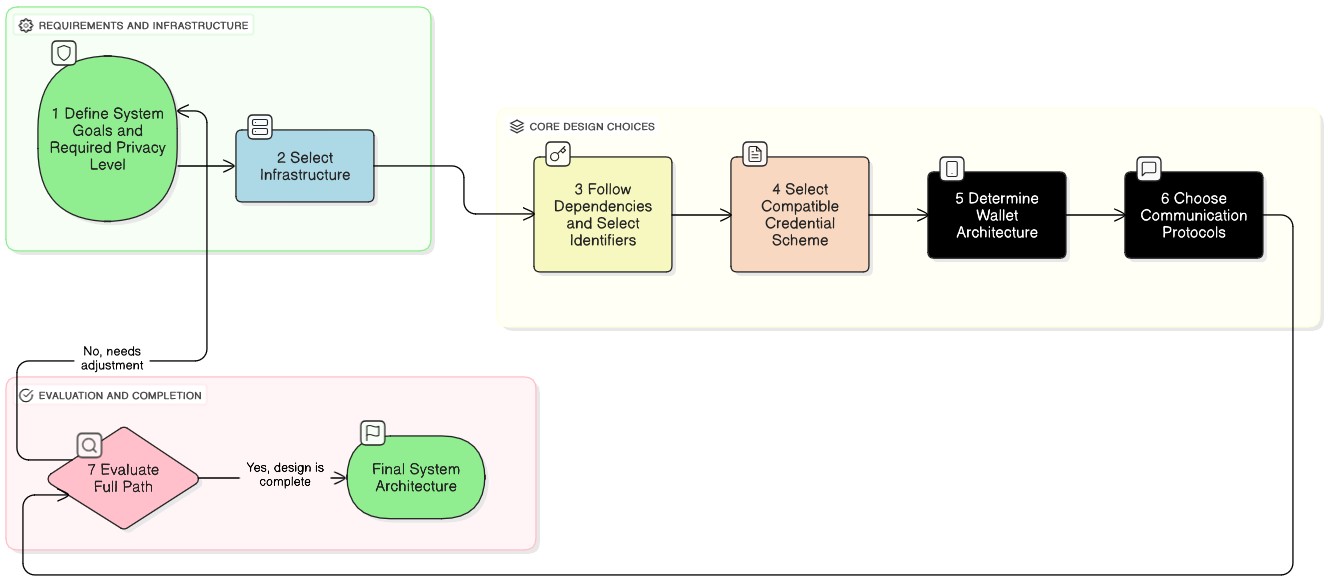}
    \caption{A step-by-step flowchart for applying DAD to SSI design.}
    \label{fig:flowchart}
\end{figure}

The step-by-step guide is presented in the flowchart in Figure \ref{fig:flowchart}. 
It identifies seven main steps, the very first being the definition of privacy goals and requirements, with the aim of analyzing and designing an SSI system.

The use cases studied for applying the guide are as follows: 

\textbf{Case A: Public University Diploma System (Path 1 in blue)} \\
1. Define Goals: the primary requirements are public verifiability and high integrity. A university is a public issuer that does not require the privacy of its identifier. Students are the holders, and their privacy is secondary to the need for employers to verify diplomas. The target is \textbf{medium privacy}. \\
2. Select Infrastructure: Given the university's public nature and existing web presence, a \textbf{web-based infrastructure} is an efficient choice. DAD shows that web-based infrastructures offer good privacy. The university can host the necessary DID document and credential schema as web-resources on their official website.\\
3. Select Identifiers: \textbf{did:web} is a web-dependent identifier that uses the domain name of the university. It is a low-privacy identifier, but it is ideal for universities as they that don't need privacy. For holders who want medium privacy and portability, a non-dependent identifier is best, so \textbf{did:key} is a great choice. \\
4. Select Credential Schema: \textbf{JWT} offers low privacy but are simple and widely compatible with the web environment. \\
5. Determine Wallet Architecture: to prioritize ease of access for students from any device, the \textbf{cloud agent} is responsible for storing credentials, thereby accepting the trade-off of reduced user-control for greater convenience. The same cloud agent can also perform DID and key generation alongside cryptographic operations. \\
6. Choose Protocols: DID-Auth and DID-Comm are the two main SSI wallet protocols. However, there are different ways to implement both protocol. For this blue path, we use \textbf{DID-auth over OIDC} to authenticate students with their university in order to request a diploma as a credential. DID-Comm is not needed since the Cloud Agent and the University/Employers can communicate using other web protocols to communicate without relying on DIDs. \\
7. Evaluate Full Path: this blue path consists of red, orange and green components. While it successfully creates a public, verifiable credential system, it offers limited privacy for the credential holder. This architecture aligns with the initial goal of prioritizing public verifiability over privacy.

\textbf{Case B: Verifiable Health Credential System (Path 2 in green)} \\
1. Define Goals: the primary requirement is maximum privacy for the holder. Unlinkability and selective disclosure of sensitive information are required, the target is \textbf{high privacy}. \\
2. Select Infrastructure: avoiding central control or surveillance requires a decentralized infrastructure. \textbf{Blockchain} with medium privacy is selected as a verifiable data registry.\\
3. Select Identifiers: a \textbf{Link Secret} offers good privacy for holders, while \textbf{did:indy} with medium privacy is ideal for issuers.\\
4. Select Credential Schema: \textbf{AnonCreds} is a standard VC schema for credentials in the Hyperledger (Indy, Aries) ecosystem. They offer good privacy and allow for ZKP and Selective Disclosure.\\
5. Determine Wallet Architecture: exclusive holder control over credentials and identifiers, as well as the rest of their health data, is required. An \textbf{Edge Agent} implementing sensitive cryptographic operations (key and identifier generation, signatures) and storing the health data locally is also required. \\
6. Choose Protocols: using a cloud agent to mediate between edge agents is an option, but peer-to-peer, end-to-end encrypted communication between edge agents is preferable. Both DID-Auth and DID-Comm offer good privacy for authentication and data exchange.\\
7. Evaluate Full Path: the resulting path consists entirely of green and orange components. It creates a system that prioritizes the privacy, control and unlinkability of the holder, thus meeting the initial goal of providing a high level of privacy.

\section{Discussion, Open Challenges and Future Work}
\label{section 8}
In this paper, we conducted an in-depth review of the SSI ecosystem and synthesized our findings into a privacy-oriented SSI layered framework. This framework is accompanied by a Design Assistance Dashboard (DAD), which is a tool that enables the quick analysis of SSI dependencies and assessment of the privacy levels of a given architecture. Unlike other models that focus on the governance or interoperability issues of SSI systems, such as ToIP, our paper presents a layered, privacy-first dissection of SSI that highlights the trade-offs often made in real-world implementations.
However, despite the promise and utility of SSI, our review points to the following open challenges:
\begin{itemize}
    \item True user sovereignty: it is difficult to achieve the SSI principles in practice. Even with a privacy-preserving design, user control is still mediated by multiple third-parties, like the wallet vendor (provider), the operating system of the device running the wallet and the device manufacturer. This can undermine user's authority over their identity, even when privacy guarantees are implemented at the SSI level.
    \item Recovery and usability: managing keys and identifiers poses a significant challenge in terms of usability. The recovery of keys and credentials is one of the largest barriers to mainstream adoption since there is a high risk of permanent loss\footnote{The Bitcoin community is familiar with this, estimates suggest that 11-18\% of the total Bitcoin supply ha been permanently lost in wallets with no access due to key loss. \url{https://www.ledger.com/academy/topics/economics-and-regulation/how-many-bitcoin-are-lost-ledger}}.
    \item Performance cost of privacy: as the credential analysis shows, stronger privacy protection comes at a higher computational cost due to larger key and signature sizes. This performance overhead makes privacy-preserving solutions impractical for large-scale applications or environments with limited resources.
\end{itemize}
In conclusion, this work provides a foundation for several future works in SSI. Our immediate goal is to continue expanding the DAD by incorporating additional components and highlighting more detailed dependencies and architectural options as new standards and components emerge. The most significant step is to develop the DAD as an interactive digital tool with a graphical user interface (GUI). This would allow users to select SSI components and choose their architecture, with the tool automatically highlighting dependencies, flagging privacy conflicts or redundancies, and potentially calculating an overall privacy score with better-defined metrics, rather than using three classes (good, medium, bad). Such a tool would enhance computer-aided design of SSI and will serve as a design and educational resource for the SSI community. Such a tool would enhance computer-aided design of SSI and serve as a valuable resource for designers and educators within the SSI community.

\section*{Funding}
This work benefited from State aid managed by the Agence Nationale de la Recherche (ANR) under the France 2030 programme, reference \textbf{ANR-22-PESN-0006} (Project TRACIA). It is also partly supported by
Agence Nationale de la Recherche under reference \textbf{ANR-23-P012-0013} (PRIMA project MoreMedDiet) 
and the chair Values and Policies of Personal Information, Institut Mines-Telecom, France. 

\section*{Declaration of Generative AI and AI-assisted technologies in the writing process}
The authors benefited from phrasing and correction suggestions provided by the free version of Writefull included within the free version of Overleaf during the writing of this paper, in order to improve readability and language of the manuscript.

\section*{Declaration of competing interest}
The authors declare that they have no known competing financial interests or personal relationships that could have appeared to influence the work reported in this paper.

\section*{Data availability}
No data was used for the research described in the article.

\newpage


\section*{Supplementary material}

**Appendix A**: *SSI breakdown by component into **wallet** and **infrastructure***

| Component | Break Down | | | |
|---|---|---|---|---|
| Identifier and Credential | Identifier | DID | | DID methods |
| | | | | DID resolvers |
| | | Link Secret | Blinded link secret | |
| | | Other | | |
| | Credential | Data model | | JSON |
| | | | | JSON-LD |
| | | Proof section | Signature scheme | JOSE |
| | | | | LDS |
| | | | | CL |
| | | | | BBS+ |
| | | | Other proof types | Timestamp |
| | | | | Proof of existence |
| | | | | other |
| | | Schema & Definition | | Published on Blockchain/DLT |
| | | | | Published on Web-based registry |
| Wallet | Hardware | Storage | Computing | Device |
| | Exchange protocol | DID-Comm | | |
| | | DID-Auth | | |
| | | Other | | |
| | Agents | Hyperledger Aries | | |
| | | uPort Wallet | | |
| | | Trinsic Wallet | | |
| | | MATTR Wallet | | |
| | | Other | | |
| Infrastructure | Blockchain, DLT... | | | |
| | Web-based | | | |
| | Hybrid (on-chain + off-chain) | | | |

**Appendix B:** A comparison between major DID methods.

| Method | Core Principle | Example Usage | DID document | DID resolution |
|---|---|---|---|---|
| did:ethr | Map an Ethereum address (which is a public key) to a DID identifier. | The Uport project | Built from smart contract events specific to a given Ethereum address. The smart contract "ethr-did-registry" is an example. | A smart contract "ethr-did-resolver" takes an DID and builds the DID document from another contract events. |
| did:web | Turns a domain name (URL) into a DID identifier. | Sidetree supports the web method, multiple other implementations and examples like Talao wallet. | The DID document is generally stored as a resource on the website itself, or in another web directory. | Resolved through the DNS since the identifier contains a website link. |
| did:key | Prefix "*did:key*" followed by a base58-btc encoded value, created from the public key. Can be generated from Ed25519, BLS, Secp, RSA... | Hyperledger Aries, Dock, multiple others | DID documents cannot be updated under this method, nor can they be deactivated. Compromised keys cannot be rotated, making it a permanent identifier. | Expanded from the DID. |
| did:indy and did:sovrin | The DID is generated from the **verkey** (a public key) associated to a private key used to sign the NYM transaction published on the ledger. The NYM transaction have 3 possible values: 0 meaning no relationship between the initial verkey and the DID, 1 means the DID is the first 16 bytes of the public key specified in the verification method in the DID document (did:sov) and 2 means the DID is derived from the verkey and the public key specified in the verification method in the DID document (the key is hashed with SHA256, truncated to 16 bytes and base58 encoded). | Hyperledger Indy, Sovrin network | The DID document is assembled and validated using transaction data found in the NYM transaction. When the verkey is null the DID is deactivated, no DID document | The resolver retrieves the NYM transaction associated to the DID from the ledger, verifies the NYM state and assembles a DID document from it following the concept specified in the previous column. |
| did:peer | 3 different methods to generate identifiers, first method generates an identifier from a single key-pair with an empty DID doc. The second one requires a genesis DID doc, and the third one is similar to the first but can include multiple key-pair for signing, authentication etc. | Peer to peer communication, decentralized networks, peer identifiers between nodes | Depending on the generation method, if it is the method "0" we don't need a DID doc since the identifier is similar to did:key, method "2" also does not require a DID doc, and in method "1" there is a genesis DID doc that registers a publishing key etc. This method does not require a DID doc in most cases since there are no endpoints to specify. | DIDs are exchanged via the Aries 0023 DID exchange protocol or sent along with signed DID doc over TLS. The DID is expanded to a DID document. |

*Appendix C:* Life cycles of different credentials

| Step | JSON JWT VC | JSON ZKP CL VC (AnonCreds) | JSON-LD LDS VC | JSON-LD BBS+ VC |
|---|---|---|---|---|
| Credential Definition | VC semantics are defined according to the IANA registry for JWT. | Credential Schema and Credential Definitions are published on a ledger. Schema and definition have unique names, which are used to retrieve them. These unique names can be found in the credential. | Annotations of JSON-LD like @context and @type refer to URLs where the schema and definitions are published. | Schema and definitions are published in web-based registries and referred to in the credential using URLs. |
| Binding the VC to the holder | The VC is bound to a DID. | The VC is anchored to a blinded version of a link secret. Credentials can also be linked to a DID of the holder or even to biometrics. | The VC is bound to a DID. | The VC is bound to a DID. It also includes a binding to a public key and doesn't reveal the DID itself. This is like link secrets found in AnonCreds. |
| Signing the VC | Attributes and their values are organized in any order in a JSON object, and entirely signed using a JWS. The result is a JWT. | The issuer generates a numeric representation and predicates of each attribute and later signs the numeric representation and the attribute with a CL signature. | The whole credential is signed with an LD signature. | The issuer signs each attribute individually using a BBS+ signature. No predicates are needed since BBS+ is malleable and the holder acts as a redactor. |
| Creating a VP from the VC | The verifier sends a NONCE to the holder to create a VP. The VC is not modifiable. It is simply encapsulated within another JWT and signed with a JWS using the holder's key associated to the DID contained in the VC and the provided NONCE to prevent replay attacks. | The verifier sends a NONCE to the holder to create a VP. The selected signed attributes and predicates from different credentials are wrapped within a VP. The proof section includes proving that the holder knows the link secret. | The verifier sends a NONCE to the holder to create a VP. The VC is not modifiable. The holder signs and puts the hole VC along with the NONCE in a VP that is signed using their keys declared in the public DID. | The verifier sends a NONCE to the holder to create a VP. The VP is created by selecting whatever attributes needed to be present from the VC. Signed attributes are sanitizable and the holder can produce ZKP attributes. |
| VP verification by the verifier | The verifier resolves both the DIDs of the issuer and of the holder. Upon finding the associated public keys, they verify the JWS found in the VC and on the VP. Further verification includes the status of the keys and of the credential itself for revocation. | The VP usually contains attributes from different credentials. The verifier must look up the credential definition that each attribute came from since the definition specifies who is the issuer and what keys are supposed to be used to verify the attribute. Credentials are bound to the same link secret, so when the holder prove they know the link secret, they assure the verifier that they are the same entity to which the attributes were issued. | The verifier resolves the DIDs of both holder and issuer and uses their public keys to verify the signatures. | If the VP is linked to a public key that does not reveal the DID, the verifier sends a challenge to the holder. If the holder can decrypt, then they prove that indeed they are the owner of that credential. The DID is not revealed. If link secrets are used, it will be the same as AnonCreds. The verification also include revocation status. |

**Appendix D:** Comparing Cloud agent and Edge agent by implemented functionalities

| Functionality | Analysis | |
|---|---|---|
| Generating DIDs and associated keys | Edge agent | The generation of keys and DIDs on the client's device is the closest to SSI requirements like ownership and control, however this comes with extra security burdens to isolate and protect these identifiers and keys from other applications installed on the edge device and from the operating system of the device. This is more secure in terms of protecting the SSI infrastructure from large scale attacks since there are no single points of failure in regards of the leakage of sensitive cryptographic material. |
| | Cloud Agent | A cloud agent operates in a less constrained environment than an edge agent. Generating cryptographic material on a cloud agent is easier, faster and more secure since they are more isolated than edge agents that heavily depend on the device they are installed on. Singing transactions and interacting online is also faster since cloud agents have better and more stable internet connection, however a cloud agent acts on behalf of multiple users and generate cryptographic material for multiple entities, making it a target for attackers and security threats and creates failure points where a data breach can prove costly for the whole SSI solution. |
| Private key storage | Edge agent | The private keys are only stored on the user's end, the user is solely responsible for the keys and their recovery in case of loss. Having the private key only on the edge agent offers users more user control and ownership and reduces chances of impersonation or fraud since no one else can sign their transactions, however, this means that they can only access the SSI system from that device and only that device, unless the private key is somehow replicated on other devices. |
| | Cloud Agent | This storage offers better recovery options and places the recovery duties on the cloud agent, proving to be more user friendly than the other choice. It can also prove more secure since the cloud agent and its administrators can establish higher security measures than normal users. In terms of accessibility, the user has more flexibility since they can access their keys and sign transactions using any device they wish since the key is on the cloud agent. |
| Signing credentials/presentations/transactions | Edge agent | The edge agent signs transactions offline and transfers them to be written on the infrastructure or to be exchanged with other entities. The signing time can be longer however, and the type and number of supported signature schemes is very limited in comparison with those a cloud agent offers. |
| | Cloud Agent | Signing time can be shorter on cloud agents offering better performance. But this signing is done on behalf of users based on their request, which opens security threats like false requests and impersonation, breaches leading to keys being stolen, etc. Cloud agents implement more various signature schemes and offer users multiple signing options. This requires trusting the cloud agent to act on the user's behalf. |
| Credential exchange | Edge agent | Edge agents can directly exchange credentials in the form of QR codes. They can also exchange these via connectivity methods like Wi-Fi, Bluetooth and NFC for close users, or via other sharing methods that are conducted outside of the SSI wallet system like social media and emails. Some cases use boot-nodes and peer-to-peer networks to exchange credentials directly without passing by a server or a cloud component. |
| | Cloud Agent | Cloud agents are used as intermediaries to relay credentials and messages between the edge agents. This exchange layer needs a level of privacy to ensure the confidentiality, integrity and authenticity. DID-Comm ensures these needs thanks to encrypted messages and protect the exchange even from the involved cloud agents that route the credentials or deliver them to target endpoints with end-to-end encryption. |
| Routing messages | Edge agent | Routing messages in a P2P network can be done via multiple algorithms, mainly |

| | | |
|---|---|---|
| | | Kademlia and others. Discovery and routing are needed to forward messages between two edge agents by passing by intermediary nodes and peers, that can be edge agents too or even nodes in the blockchain infrastructure if one is used in the SSI solution. P2P networks are generally open, and the barrier of entry is very low, meaning that it is a trustless environment, so extra security measures need to be taken for routing confidential data like messages between edge agents. When an edge device is not available for message delivery or the endpoint is not always kept alive, messages might have to be stored on other nodes until they reach their receiver, which also can cause security and privacy threats. |
| | Cloud Agent | In many SSI solutions and designs, the messages are enveloped by the edge agent and transferred to a cloud agent or a mediator to be routed from one agent to another until it reaches the endpoint or the edge agent. Same as credential exchange the privacy of routing messages between agents can be guaranteed to an extent since cloud agents can still know – to an extent – the receiver and the sender. Cloud agents however are generally trusted entities and are subject to control and supervision by legislation, unlike nodes in a P2P network. |
| Storing messages/credentials | Edge agent | This makes them private and less likely to be stolen in cases of data breaches that target online cloud storage. This also inspires more control and ownership since they are hosted on the user's device and not hosted anywhere else. However, extra measures to isolate them and secure them from other applications present on the device need to be taken. Moreover, they can only be accessible through that device only, and recovery and backup measures must be put in place. |
| | Cloud Agent | This makes credentials and messages accessible from any device as soon as the user connects to the cloud agent. The messages and credentials are generally encrypted before being stored on the cloud agent or even encrypted by the cloud agent itself to protect them from any possible threats. This undermines the ownership and control that SSI requires to facilitate SSI interactions and offer users an easier more portable way to present their credentials and communicate. |

**_Appendix E:_** Comparing on-chain and off-chain SSI components

| | Privacy and Security remarks |
|---|---|
| On-chain DIDs | A DID published on the ledger is a public DID and generally intended for public use. The DID is immutable and always available for resolution by any entity, even curious entities. |
| Off-chain DIDs | The DID is either published in a public internet-based registry or kept by the user that chooses to share it with whom they want. These DIDs can be public or private, and the private ones are ideal for privacy preserving communication. |
| On-chain DID Documents | An on-chain DID is transparent and public, anyone can see it and any updates or modifications need to be published, augmenting the traceability of DID owner activities.<br>The DID document integrity and authenticity are guaranteed.<br>Old versions of the DID document are permanent and cannot be removed from the ledger, which can pose problems if they contain sensitive data. |
| Off-chain DID documents | Accessing off-chain DID documents for resolution is easier to control off-chain since we can implement access control policies to better control who can see the data.<br>DID owners must sign the DID documents for integrity and authenticity, especially after updates. |
| Blockchain-based resolver (via smart contract and smart contract events) | The resolver assembles the DID document data from smart contract and blockchain events, these can be public or private depending on the smart contract and the ledger configuration. In the case of public blockchains with public smart contracts – like Ethereum UPort – anyone can resolve a DID. The security of the resolution is guaranteed since it is based on smart contracts and subject to consensus, the resolver will correctly assemble the document and the verification of the output is easy. |
| Web-based resolver | A web-based resolver acts as an API to retrieve a DID document that matches a DID from an internet-based registry. The privacy of the resolution can be low in the case of DNS based DIDs like **did:web** and **did:dns**, and even in other cases where the resolver is a third-party software that the wallet agents use to resolve and communicate between themselves. Security too must be addressed as we can have cases of DNS poisoning or MIM attacks during the resolution. |
| On-chain credential proofs | Credential anchoring or on-chain proofs can range from full proofs of the credential to only proof of existence of the credential without revealing the content. Anchoring a credential onto the ledger – in the form of a published hash – is solid proof that an issuer created a credential for a holder. Generally, the transaction will include DIDs from issuer and holder, and a hash of the credential to ensure integrity. Blockchain timestamp of the transaction serves as credential creation date, and revocation can be done by issuing a new transaction with the same credential hash indicating it was revoked. The permanent hash on the ledger can turn into a bad privacy-choice as soon as hash dictionaries and computing power advance enough to jeopardize the privacy of the hash. The on-chain proofs are as secure as the transaction that was made to create them, meaning they benefit from the blockchain security but rely on the security of the publishing entity. Anchored credentials lack flexibility and privacy enhancing methods like selective disclosure. The only privacy here is a proof of existence without revealing the content, which is a basic form of Zero-Knowledge proofs. |
| Off-chain credential proofs | These are mainly proofs that are independent of the blockchain network like digital signatures and electronic seals and any other proofing method the issuer can apply on a VC to prove the integrity, authenticity and bind it to the user for ownership proofing. Digital signatures offer higher flexibility than published hashes and offer privacy-preserving methods like selective disclosure and ZKP. |


\label{appendix:E}

\end{document}